\documentclass{article}
\usepackage{latexsym}

\def\dag{\dagger} \def\pd{\partial} \def\pp{\prime} \def\a{\alpha} \def\b{\beta} \def\dl{\delta} \def\s{\sigma} \def\vphi{\varphi} \def\eps{\epsilon} \def\veps{\varepsilon} \def\lam{\lambda}  \def\gm{\gamma}  \def\om{\omega}  \def\tkappa{{\tilde \kappa}}
\def\sq{\sqrt} \def\fr{\frac} \def\half{\frac{1}{2}}
\def\rb{{\rm b}} \def\rc{{\rm c}} \def\gh{{\rm gh}}  \def\BRST{{\rm BRST}} 
 
\def\hg{{\hat g}} \def\bg{{\bar g}}  \def\nb{\nabla} \def\hnb{{\hat \nabla}} \def\bnb{{\bar \nabla}} \def\hDelta{{\hat \Delta}}  \def\hR{{\hat R}} \def\bR{{\bar R}} \def\hE{{\hat E}}    

\def\calA{{\cal A}} \def\calB{{\cal B}} \def\calC{{\cal C}}
\def\bx{{\bf x}} \def\by{{\bf y}} \def\bk{{\bf k}}
 \def\P{{\rm P}} 
  
\def\h{{\sf h}} \def\u{{\sf u}} \def\P{{\sf P}} \def\sfc{{\sf c}} \def\sfd{{\sf d}} \def\sfe{{\sf e}}

\def\lap3{~| \!\!\! \partial^2} \def\dlap3{~| \!\!\! \partial^4} \def\invlap3{~| \!\!\! \partial^{-2}}
\def\l:{: \!}\def\r:{\! :}
\def\lang{\langle} \def\rang{\rangle}
\def\d3x{d^3 {\bf x}~}

\begin{document}

\begin{titlepage}

\begin{flushright}
{\sc September 2011}
\end{flushright}

\vspace{5mm}

\begin{center}
{\large {\bf Background Free Quantum Gravity based on Conformal Gravity and Conformal Field Theory on $M^4$}} 
\end{center}

\vspace{5mm}

\begin{center}
{\sc Ken-ji Hamada}\footnote{E-mail address: hamada@post.kek.jp; URL: http://research.kek.jp/people/hamada/}
\end{center}

\begin{center}
{\it Institute of Particle and Nuclear Studies, KEK, Tsukuba 305-0801, Japan} \\ and \\
{\it Department of Particle and Nuclear Physics, The Graduate University for Advanced Studies (Sokendai), Tsukuba 305-0801, Japan}
\end{center}

\begin{abstract}
We study four dimensional quantum gravity formulated as a certain conformal field theory at the ultraviolet fixed point, whose dynamics is described by the combined system of Riegert-Wess-Zumino and Weyl actions. Background free nature comes out as quantum diffeomorphism symmetry by quantizing the conformal factor of the metric field nonperturbatively. In this paper, Minkowski background $M^4$ is employed in practice. The generator of quantum diffeomorphism that forms conformal algebra is constructed. Using it, we study the composite scalar operator that becomes a good conformal field. We find that physical fields are described by such scalar fields with conformal dimension 4. Consequently, tensor fields outside the unitarity bound are excluded. Computations of quantum algebra on $M^4$ are carried out in the coordinate space using operator products of the fields. The nilpotent BRST operator is also constructed.  
\end{abstract}

\end{titlepage}


\section{Introduction}
\setcounter{equation}{0}
\noindent

To break the wall of the Planck scale, the background free nature of spacetime will be necessary. Indeed, at the epoch beyond the Planck energy scale, spacetime fluctuations will be so great and geometry loses its classical meaning. It simply means that there is no fixed scale and no special point in space. Therefore, the problem that a particle excitation with the mass over Planck scale becomes a black hole\footnote{ 
The Compton wave length, which represents a typical size of particle, becomes smaller than the Schwarzschild radius, and thus its information will be lost.
} 
can be avoided basically because such a standard description of particles propagating on a classical background itself breaks down. Thus, the background free nature will throw light on the unitarity problem in gravity.

To be background free, gravity should be quantized nonperturbatively. Two dimensional quantum gravity, known as the Liouville theory, indeed realizes such a background free picture as conformal symmetry by quantizing the conformal factor of the metric field exactly \cite{polyakov, kpz, dk}. The four dimensional counter model is known to be described by the Riegert-Wess-Zumino action \cite{riegert}, which is quantized in \cite{am, amm92, amm97, hs, hamada02, hh, hamada05, hamada09a, hamada09b, hamada11, nova}. The background free model we will study is given by the ultraviolet (UV) limit of renormalizable quantum gravity that is formulated on the basis of conformal gravity incorporating this action systematically \cite{hamada02, hamada09b, nova}.

The model is characterized by how to decompose the metric field into the conformal factor $e^{2\phi}$ and the traceless tensor field $h_{\mu\nu}$, where we call $\phi$ the Riegert field. The former is quantized nonperturbatively without introducing its own coupling constant, while the latter is handled by the perturbation theory as
\begin{equation}
     g_{\mu\nu}=e^{2\phi}(\hg e^{th})_{\mu\nu}= e^{2\phi} \left( \hg_{\mu\nu} + t h_{\mu\nu} + \cdots \right) ,
       \label{metric decomposition}
\end{equation}
where $tr(h)=h^\mu_{~\mu}=0$ and $\hg_{\mu\nu}$ is the background metric. $t$ is the dimensionless coupling constant indicating asymptotic freedom \cite{hamada02, hamada09b}.\footnote{ 
This perturbation theory is characterized by the condition that the Riegert field $\phi$ does not receive renormalization at all orders of the perturbation.
} 
The expansion by $t$ represents the perturbation from quantum spacetime fully fluctuating under the restriction that the Weyl tensor vanishes.

We here consider the model at the UV fixed point of $t = 0$, where exact conformal symmetry comes out as a part of quantum diffeomorphism invariance, namely, background metric independence, realized by taking the sum over all possible conformally flat configurations.\footnote{ 
It should be distinguished from the invariance under the Weyl rescaling. Throughout our works, we respect diffeomorphism invariance. Therefore, conformal anomalies \cite{cd, ddi, duff} appear as indispensable elements to preserve diffeomorphism invariance. The coupling $t$ measures a degree of deviation from conformal symmetry. With the increase of it, spacetime will make a transition from quantum phase to classical Einstein phase dynamically \cite{hy, hhy06, hhsy, hhy10, nova}.
} 

The background metric can be chosen arbitrary so long as it is conformally flat. Indeed, in the previous papers \cite{amm97, hh, hamada05, hamada09a, hamada11}, the model have been studied on the curved background $R \times S^3$, which has various advantages such that structure of conformal algebra and analysis of physical states become similar to the cases of the Virasoro algebra on $R \times S^1$. In this paper, we reconsider the model by employing the Minkowski background $M^4$. The advantages of using this background are that it is familiar to most of the researchers in this area and physical field operators become easier to handle compared with those defined on the curved background.

The aim of these studies is to clarify physical properties of the field from the viewpoint of symmetry. The fact that conformal symmetry originates from diffeomorphism symmetry, namely, gauge symmetry, is significant for unitarity, because it means that conformally variant fields are excluded from physical fields. Consequently, physical fields are described as spacetime volume integrals of scalar fields with conformal dimension 4, while fields with tensor indices are excluded.\footnote{ 
This result is consistent with astrophysical observations that scalar fluctuations with power-law spectra are dominant in the early universe.
}  

This paper is organized as follows. We briefly summarize the model in the next section. In section 3, we quantize the Riegert-Wess-Zumino action and construct the generator of conformal symmetry on $M^4$. In section 4, we study the transformation laws of the Riegert field and its exponential function, and then discuss the condition for physical fields. The proof that the generator indeed forms conformal algebra at the quantum level is given in section 5. The Weyl action is quantized in section 6. The generator of conformal symmetry and the transformation laws of traceless tensor fields are studied there. In section 7, we study quantum diffeomorphism symmetry in the context of the Becchi-Rouet-Stora-Tyupin (BRST) formalism \cite{brs, ko, kato}. The nilpotent BRST operator is constructed and physical field conditions are rewritten in terms of the BRST operator. Section 8 is devoted to conclusion.

\section{The Model}
\setcounter{equation}{0}
\noindent

In this section, we briefly review the model of four dimensional quantum gravity we will study. The action and the gauge symmetry of the model are summarized.

\subsection{The Action}
\noindent

The perturbation theory about configurations with vanishing Weyl tensor (\ref{metric decomposition}) is defined by the Weyl action divided by the square of the coupling constant as $I=-(1/t^2) \int d^4 x \sq{-g}C_{\mu\nu\lam\s}^2$. The path integral over the metric field is described as
\begin{equation}
    Z = \int [dg]_g \exp(iI)=\int [d\phi dh ]_\hg \exp ( iS+iI ).
\end{equation} 
Here, $S$ is the Wess-Zumino action induced from diffeomorphism invariant measure $[dg]_g$. It is necessary to preserve diffeomorphism invariance when we rewrite the path integral using the practical measure defined on the background metric $[d\phi dh]_\hg$.

We here consider the UV limit of $t \to 0$ indicated by the asymptotically free behavior of the coupling \cite{hamada02, hamada09b}. The Weyl action then becomes quadratic in the traceless tensor field and the induced action $S$ is given by the Riegert-Wess-Zumino action \cite{riegert},
\begin{equation}
     S_{\rm RWZ} = -\fr{b_1}{(4\pi)^2} \int d^4 x \sq{-\hg} \left\{ 2\phi \hDelta_4 \phi 
                   + \left( \hat{G}_4 -\fr{2}{3} \hnb^2 \hR \right) \phi \right\} ,
             \label{Riegert action}
\end{equation}
where $G_4$ is the Euler density and $\sq{-g}\Delta_4$ is the conformally invariant fourth-order differential operator for a scalar variable defined by $\Delta_4=\nb^4 +2R^{\mu\nu}\nb_\mu \nb_\nu -2R\nb^2/3 + \nb^\mu R \nb_\mu/3$ and $\nb^2=\nb^\mu \nb_\mu$. The coefficient $b_1$ has the physically right sign of positive.\footnote{ 
It has been computed to be $b_1 = ( N_X + 11N_W/2 + 62 N_A )/360 + 769/180$, where $N_X$, $N_W$, and $N_A$ are numbers of scalar fields, Weyl fermions, and gauge fields, respectively \cite{cd, ddi, duff}, and the last term is the loop correction obtained by quantizing the Riegert-Wess-Zumino and Weyl actions \cite{amm92, hs}.
} 
This action is the four dimensional analog of the Liouville action in two dimensional quantum gravity.

In the following, the background metric $\hg_{\mu\nu}$ is taken to be the Minkowski metric $\eta_{\mu\nu}=(-1,1,1,1)$. The spacetime coordinate is described by $x^\mu=(\eta, x^i)$ and $x^2=x_\mu x^\mu =-\eta^2 + \bx^2$. The d'Alembertian is denoted by $\pd^2= \pd^\lam \pd_\lam=-\pd_\eta^2 + \lap3$, where $\lap3=\pd_i \pd^i$ is the Laplacian of three dimensional space.

\subsection{Gauge Symmetry}
\noindent

Under the metric decomposition (\ref{metric decomposition}), diffeomorphism defined by $\dl_\xi g_{\mu\nu}=g_{\mu\lam}\nb_\nu \xi^\lam + g_{\nu\lam}\nb_\mu \xi^\lam$ is decomposed into the transformations of Riegert and traceless tensor fields as
\begin{eqnarray}
   \dl_\xi \phi &=& \xi^\lam \pd_\lam \phi  
                       + \fr{1}{4} \pd_\lam \xi^\lam ,     
            \nonumber \\
   \dl_\xi h_{\mu\nu} &=& \fr{1}{t} \left( \pd_\mu \xi_\nu +\pd_\nu \xi_\mu
           - \half \eta_{\mu\nu} \pd_\lam \xi^\lam \right)
           + \xi^\lam \pd_\lam h_{\mu\nu}   
                \nonumber    \\ 
    &&     + \half h_{\mu\lam} \left( \pd_\nu \xi^\lam 
                  - \pd^\lam \xi_\nu \right) 
           + \half h_{\nu\lam} \left( \pd_\mu \xi^\lam 
                  - \pd^\lam \xi_\mu \right) 
           + o(t\xi h^2),
               \label{diffeomorphism for traceless mode}
\end{eqnarray}
where $\xi_\mu =\eta_{\mu\nu} \xi^\nu$.

There are two types of diffeomorphism symmetry at the vanishing coupling limit. The first is the well-known gauge symmetry that the kinetic term of the Weyl action has. Introducing the gauge parameter $\kappa^\mu=\xi^\mu/t$ and taking the limit $t \to 0$ with $\kappa^\mu$ fixed, the diffeomorphism is expressed as 
\begin{eqnarray}
      \dl_\kappa h_{\mu\nu} = \pd_\mu \kappa_\nu + \pd_\nu \kappa_\mu - \half \eta_{\mu\nu} \pd_\lam \kappa^\lam 
               \label{gauge transformation}
\end{eqnarray}
and $\dl_\kappa \phi =0$. This gauge symmetry will be fixed completely by taking the radiation gauge, which will be discussed when the Weyl action is quantized in section 6.

The second is the conformal symmetry we will study in this paper. It is the residual diffeomorphism symmetry in the radiation gauge with a gauge parameter $\xi^\mu =\zeta^\mu$ satisfying the conformal Killing equation,
\begin{equation}
    \pd_\mu \zeta_\nu + \pd_\nu \zeta_\mu 
                 - \fr{1}{2} \eta_{\mu\nu} \pd_\lam \zeta^\lam =0.
           \label{conformal Killing equation}
\end{equation}
Since the lowest term of the transformation of $h_{\mu\nu}$ (\ref{diffeomorphism for traceless mode}) vanishes in this case, the second term becomes effective. Thus, we obtain \cite{hamada09a}
\begin{eqnarray}
    \dl_\zeta \phi &=& \zeta^\lam \pd_\lam \phi + \fr{1}{4} \pd_\lam \zeta^\lam 
           \nonumber \\
    \dl_\zeta h_{\mu\nu} &=& \zeta^\lam \pd_\lam h_{\mu\nu} 
              + \half h_{\mu\lam} \left( \pd_\nu \zeta^\lam - \pd^\lam \zeta_\nu \right)
              + \half h_{\nu\lam} \left( \pd_\mu \zeta^\lam - \pd^\lam \zeta_\mu \right) .
         \label{conformal transformation}
\end{eqnarray}
The kinetic term of the Weyl action becomes invariant under this transformation without taking into account self-interaction terms. The transformation $\dl_\zeta$ is a conformal transformation considering quantum gravity as a quantum field theory on $M^4$.

Although the residual gauge degrees of freedom are finite, the gauge symmetry $\dl_\zeta$ is quite strong because the right-hand sides of (\ref{conformal transformation}) are field-dependent so that the transformations mix all modes in the fields. Thus, these modes themselves are not gauge invariant, including ghost modes.

Here, we emphasize that the shift term in $\dl_\zeta \phi$ represents that this transformation is of diffeomorphism origin and the Riegert field is not a simple dimensionless scalar. This gauge symmetry leads to the invariance under the conformal change of the background metric as $\eta_{\mu\nu} \to (1+ \pd_\lam \zeta^\lam/2)\eta_{\mu\nu}$ (see also Appendix A). The background free nature is thus realized as gauge symmetry quantum mechanically.

The generator of conformal transformation $\dl_\zeta$ is defined by using the stress tensor $T_{\mu\nu}$ satisfying the traceless and conservation conditions as
\begin{equation}
     Q_\zeta = \int d^3 \bx \zeta^\lam T_{\lam 0} ,
         \label{definition of Q_zeta}
\end{equation}
which is conserved such that $\pd_\eta Q_\zeta =0$.

The conformal Killing vectors for translations, Lorentz transformations, dilatations and special conformal transformations denoted by $\zeta^\lam_{T,L,D,S}$ are given by
\begin{eqnarray}
   (\zeta_T^\lam)_\mu &=& \dl^\lam_{~\mu} ,  
          \nonumber \\
   (\zeta_L^\lam)_{\mu\nu} &=& x_\mu \dl^\lam_{~\nu} - x_\nu \dl^\lam_{~\mu} ,
          \nonumber \\
   \zeta_D^\lam &=& x^\lam , 
         \nonumber \\
   (\zeta_S^\lam)_\mu &=& x^2 \dl^\lam_{~\mu} -2 x_\mu x^\lam .
          \label{conformal Killing vector}
\end{eqnarray}
Substituting these 15 vectors into (\ref{definition of Q_zeta}), we obtain the generators denoted as follows: $P_\mu$ of translations, $M_{\mu\nu}$ of Lorentz transformations, $D$ of dilatations and $K_\mu$ of special conformal transformations, 
\begin{eqnarray}
     P_\mu &=& \int d^3 \bx T_{\mu 0} ,
         \nonumber \\
     M_{\mu\nu} &=& \int d^3 \bx \left( x_\mu T_{\nu 0} - x_\nu T_{\mu 0} \right) ,
         \nonumber \\
      D &=& \int d^3 \bx x^\lam T_{\lam 0} , 
         \nonumber \\
      K_\mu &=& \int d^3 \bx \left( x^2 T_{\mu 0} -2 x_\mu x^\lam T_{\lam 0} \right)  .
          \label{definition of generator}
\end{eqnarray}
These generators form the closed algebra of conformal symmetry.

In the following sections, we explicitly construct the generators of the transformations $\dl_\zeta \phi$ and $\dl_\zeta h_{\mu\nu}$ from the Riegert-Wess-Zumino and Weyl actions, respectively. The shift term in $\dl_\zeta \phi$ is just generated from the linear term in the Riegert-Wess-Zumino action.

\section{Riegert Field}
\setcounter{equation}{0}
\noindent

In this section, we quantize the Riegert field on $M^4$ and then derive the generator of the transformation $\dl_\zeta \phi$.

\subsection{Canonical Quantization}
\noindent

The Riegert-Wess-Zumino action on $M^4$ is given by $-(b_1/8\pi^2) \int d^4 x \phi \pd^4 \phi$. Here, the second term in the action (\ref{Riegert action}) vanishes, but it gives crucial contributions to the stress tensor given later.

The Riegert field is quantized by introducing new variable
\begin{equation}
    \chi = \pd_\eta \phi .
        \label{variable chi}
\end{equation}
The Riegert-Wess-Zumino action is then written in the second order form as
\begin{eqnarray}
    S_{\rm RWZ} = \int d^4 x \left\{ 
                    -\fr{b_1}{8\pi^2} \left[ \left( \pd_\eta \chi \right)^2 + 2 \chi \lap3 \chi 
                     + \left( \lap3 \phi \right)^2 \right] + v \left( \pd_\eta \phi - \chi \right) 
                    \right\} ,
             \label{Riegert second order action}
\end{eqnarray}
where $v$ is Lagrange multiplier.

Following the Dirac's procedure \cite{dirac}, we remove the variable $v$ and its conjugate momentum by solving the constraints.\footnote{ 
The constraints are given by $\vphi_1= \P_\phi- v \simeq 0$ and $\vphi_2 = \P_v \simeq 0$. The Dirac bracket is defined by $\{ F,G \}_D = \{F,G \}_P - \{ F,\vphi_a \}_P C^{-1}_{ab} \{ \vphi_b, G \}_P $, where $C_{ab}=\{ \vphi_a,\vphi_b \}_P$. The equation of motion in the constraint system is given by $\{F, H \}_D =\pd_\eta F$. The quantization is completed by replacing the Dirac bracket with the commutator multiplied by the factor $-i$. 
} 
The phase space then reduces to the submanifold spanned by four canonical variables $\chi$, $\phi$, and their conjugate momenta defined by 
\begin{eqnarray}
     \P_\chi &=& - \fr{b_1}{4\pi^2} \pd_\eta \chi, 
                \nonumber \\
     \P_\phi &=& - \pd_\eta \P_\chi - \fr{b_1}{2\pi^2} \lap3 \chi .
         \label{momentum of Riegert field}
\end{eqnarray}
The canonical commutation relations are set as
\begin{equation}
     \left[ \phi(\eta, \bx), \P_\phi(\eta, \bx^\pp) \right] 
       = \left[ \chi(\eta, \bx), \P_\chi(\eta, \bx^\pp) \right] = i \dl_3 (\bx -\bx^\pp),
           \label{canonical commutation relation}
\end{equation}
and otherwise vanishes. The Hamiltonian is given by
\begin{equation}
    H = \int d^3 \bx \l: \left\{ -\fr{2\pi^2}{b_1} \P_\chi^2 + \P_\phi \chi 
                    + \fr{b_1}{8\pi^2} \left[ 2 \chi \lap3 \chi + \left( \lap3 \phi \right)^2 \right] 
                       \right\} \r: ,
\end{equation}
where $\l:~~ \r:$ denotes the normal ordering.

The equation of motion of the Riegert field is given by $\pd^4 \phi=0$, which can be expressed in terms of the momentum variable as $\pd_\eta \P_\phi = -(b_1/4\pi^2) \dlap3 \phi$. The solutions are given by $e^{ik_\mu x^\mu}$ and $\eta e^{ik_\mu x^\mu}$ and their complex conjugates, where $k_\mu x^\mu= -\om \eta + \bk \cdot \bx$ and $\om=|\bk|$.

The Riegert field is expanded by using these solutions, which is decomposed into the annihilation and creation parts as $\phi = \phi_< + \phi_>$, where $\phi_> = \phi_<^\dag$ and
\begin{equation}
    \phi_<(x) = \fr{\pi}{\sq{b_1}} \int \fr{d^3 \bk}{(2\pi)^{3/2}} \fr{1}{\om^{3/2}}  
                    \left\{ a(\bk) + i\om \eta b(\bk) \right\} e^{ik_\mu x^\mu} .
\end{equation}
By substituting this into expressions (\ref{variable chi}) and (\ref{momentum of Riegert field}), we obtain the annihilation parts of other field variables, 
\begin{eqnarray}
    \chi_< (x) &=& -i \fr{\pi}{\sq{b_1}} \int \fr{d^3 \bk}{(2\pi)^{3/2}} \fr{1}{\om^{1/2}}
                     \left\{ a(\bk) +( -1 +i\om \eta) b(\bk) \right\} e^{ik_\mu x^\mu} ,
               \nonumber \\
    \P_{\chi <} (x) &=& \fr{\sq{b_1}}{4\pi} \int \fr{d^3 \bk}{(2\pi)^{3/2}} \om^{1/2}
                     \left\{ a(\bk) +( -2 +i\om \eta) b(\bk) \right\} e^{ik_\mu x^\mu} ,
               \nonumber \\
    \P_{\phi <} (x) &=& -i \fr{\sq{b_1}}{4\pi} \int \fr{d^3 \bk}{(2\pi)^{3/2}} \om^{3/2}
                    \left\{ a(\bk) +( 1 +i\om \eta) b(\bk) \right\} e^{ik_\mu x^\mu} 
             \label{expansion of chi and momentum variables}
\end{eqnarray}
such that $\chi=\chi_< + \chi_>$ with $\chi_>=\chi_<^\dag$ and so on. The commutation relations for mode operators are then given by the form, 
\begin{eqnarray}
      \left[ a(\bk), a^\dag(\bk^\pp) \right] &=& \dl_3 (\bk -\bk^\pp), 
                \nonumber \\
      \left[ a(\bk), b^\dag(\bk^\pp) \right] &=& \left[ b(\bk), a^\dag(\bk^\pp) \right] = \dl_3 (\bk -\bk^\pp), 
                \nonumber \\
      \left[ b(\bk), b^\dag(\bk^\pp) \right] &=& 0 .
\end{eqnarray}
The Hamiltonian is given by
\begin{equation}
     H = \int d^3 \bk  \om \left\{ a^\dag(\bk) b(\bk) + b^\dag(\bk) a(\bk) - 2 b^\dag(\bk) b(\bk) \right\} .
\end{equation}

\subsection{Generators of Conformal Symmetry}
\noindent

The stress tensor of the Riegert field is given by
\begin{eqnarray}
  T_{\mu\nu} &=& -\fr{b_1}{8\pi^2} \biggl\{
               -4 \pd^2 \phi \pd_\mu \pd_\nu \phi    + 2 \pd_\mu \pd^2 \phi \pd_\nu \phi 
               + 2 \pd_\nu \pd^2 \phi \pd_\mu \phi   +\fr{8}{3} \pd_\mu \pd_\lam \phi \pd_\nu \pd^\lam \phi 
                   \nonumber \\
            && -\fr{4}{3} \pd_\mu \pd_\nu \pd_\lam \phi \pd^\lam \phi  
               +\eta_{\mu\nu} \left( 
                   \pd^2 \phi \pd^2 \phi   -\fr{2}{3} \pd^2 \pd^\lam \phi \pd_\lam \phi
                   -\fr{2}{3} \pd_\lam \pd_\s \phi \pd^\lam \pd^\s \phi 
               \right) 
                   \nonumber \\
            && -\fr{2}{3} \pd_\mu \pd_\nu \pd^2 \phi  + \fr{2}{3} \eta_{\mu\nu} \pd^4 \phi 
             \biggr\} .
\end{eqnarray}
Here, the last two linear terms come from the variation of the second term in the action (\ref{Riegert action}). The traceless and conservation conditions are satisfied by using the equation of motion as $T^\lam_{~\lam} = -(b_1/4\pi^2) \pd^4 \phi =0$ and $\pd^\mu T_{\mu\nu} = -(b_1/4\pi^2) \pd^4 \phi \pd_\nu \phi =0$, respectively.

In terms of four canonical variables, the $(00)$ component is written as
\begin{eqnarray}
   T_{00} &=& -\fr{2\pi^2}{b_1} \P_\chi^2    + \P_\phi \chi    - \P_\chi \lap3 \phi  - \pd_k \P_\chi \pd^k \phi
                    \nonumber \\
         && + \fr{b_1}{8\pi^2} \left(
                \fr{2}{3} \chi \lap3 \chi   -\fr{4}{3} \pd_k \chi \pd^k \chi  + \lap3 \phi \lap3 \phi
                - \fr{2}{3} \pd_k \lap3 \phi \pd^k \phi      -\fr{2}{3} \pd_k \pd_l \phi \pd^k \pd^l \phi
                \right)
                    \nonumber \\
         && +\fr{1}{3} \lap3 \P_\chi   + \fr{b_1}{12\pi^2} \dlap3 \phi
\end{eqnarray}
and the $(0j)$ component is 
\begin{eqnarray}
   T_{0j} &=& \fr{2}{3} \P_\chi \pd_j \chi      - \fr{1}{3} \pd_j \P_\chi \chi    + \P_\phi \pd_j \phi
                 \nonumber \\
          && + \fr{b_1}{8\pi^2} \left( 
                  4 \pd_j \chi \lap3 \phi    -\fr{8}{3} \pd_k \chi \pd_j \pd^k \phi 
                  - 2 \chi \pd_j \lap3 \phi   + 2 \lap3 \chi \pd_j \phi 
                  + \fr{4}{3} \pd_j \pd_k \chi \pd^k \phi   \right)
                 \nonumber \\
          && - \fr{1}{3} \pd_j \P_\phi     - \fr{b_1}{12\pi^2} \pd_j \lap3 \chi .
\end{eqnarray}

Substituting these components of stress tensor into (\ref{definition of generator}), we obtain the generators of translations as 
\begin{eqnarray}
     P_0 &=& H = \int d^3 \bx \calA , 
            \nonumber \\
     P_j &=& \int d^3 \bx \calB_j ,
\end{eqnarray}
where the local operators $\calA$ and $\calB_j$ are defined by
\begin{eqnarray}
    \calA &=& -\fr{2\pi^2}{b_1} \l: \P_\chi^2 \r: + \l: \P_\phi \chi \r: 
              + \fr{b_1}{8\pi^2} \left( 2 \l: \chi \lap3 \chi \r: + \l: \lap3 \phi \lap3 \phi \r: \right) ,
             \nonumber \\
    \calB_j &=& \l: \P_\chi \pd_j \chi \r:  + \l: \P_\phi \pd_j \phi \r: .           
\end{eqnarray}
Here, the normal ordering is taken. The generators of Lorentz transformations are given by\footnote{ 
Note that although $M_{0j}$ and also $D$ and $K_\mu$ have explicit dependence on the time coordinate $\eta$, these generators are indeed conserved such that $\pd_\eta M_{0j}=\pd_\eta D= \pd_\eta K_\mu=0$. 
} 
\begin{eqnarray}
     M_{0j} &=& \int d^3 \bx \left\{ -\eta \calB_j - x_j \calA - \l: \P_\chi \pd_j \phi \r: \right\} ,
              \nonumber \\
     M_{ij} &=& \int d^3 \bx \left\{  x_i \calB_j -x_j \calB_i \right\} .
\end{eqnarray}
The generators $P_\mu$ and $M_{\mu\nu}$ form the Poincar\'{e} algebra.

The generators of dilatations and special conformal transformations are given by
\begin{equation}
     D = \int d^3 \bx \left\{ \eta \calA + x^k \calB_k  + \l: \P_\chi \chi \r: + \P_\phi \right\} 
\end{equation}
and
\begin{eqnarray}
    K_0 &=& \int d^3 \bx \biggl\{ 
               \left( \eta^2 + \bx^2 \right) \calA + 2\eta x^k \calB_k 
               + 2\eta \l: \P_\chi \chi \r:   + 2 x^k \l: \P_\chi \pd_k \phi \r: 
                 \nonumber \\
        && \qquad\qquad 
              - \fr{b_1}{4\pi^2} \left( 2 \l: \chi^2 \r: + \l: \pd_k \phi \pd^k \phi \r: \right)
              + 2 \eta \P_\phi  + 2 \P_\chi  \biggr\} ,
                \nonumber \\
   K_j &=& \int d^3 \bx \biggl\{
               \left( -\eta^2 + \bx^2 \right) \calB_j  -2 x_j x^k \calB_k
               - 2\eta x_j \calA     - 2x_j  \l: \P_\chi \chi \r:  
                  \nonumber \\
       && \qquad\qquad 
               - 2\eta \l: \P_\chi \pd_j \phi \r: 
                 -\fr{b_1}{2\pi^2} \l: \chi \pd_j \phi \r: - 2x_j \P_\phi
            \biggr\} .
\end{eqnarray}
The generators $D$ and $K_\mu$ have the linear terms that generate the shift term of the transformation $\dl_\zeta \phi$.

\section{Conformal Fields and Physical Fields}
\setcounter{equation}{0}
\noindent

We first study the transformation properties of the Riegert field and its exponential function using calculation techniques based on the operator product expansion (OPE) in the coordinate space. We then discuss the condition for physical fields.

\subsection{Conformal Fields}
\noindent

Consider the operator product among the hermitian operators, and only this case is considered. The operator product of two hermitian operators $A$ and $B$ is defined by 
\begin{equation}
    A(x)B(y) = \lang A(x)B(y) \rang + \l: A(x)B(y) \r:.
\end{equation}
The singular part, or the two-point correlation function, is given by
\begin{equation}
     \lang A(x)B(y) \rang = \left[ A_<(x), B_>(y) \right] ,
\end{equation}
where $A_<$ is the annihilation part of $A$ and $B_>$ is the creation part of $B$ as defined before. Since $\l: A(x)B(y) \r:= \l: B(y)A(x) \r:$, the commutator of $A$ and $B$ can be expressed as
\begin{equation}
    \left[ A(x), B(y) \right] = \lang A(x)B(y) \rang - \lang B(y)A(x) \rang .
\end{equation}
Since the second term in the right-hand side is the hermitian conjugate of the first one such as $\lang B(y)A(x) \rang = \lang A(x)B(y) \rang^\dag$, the commutator vanishes if it is a real function.

Let us calculate the operator product of the Riegert field. The singular part is computed as
\begin{eqnarray}
    \lang \phi(x) \phi(x^\pp) \rang 
          &=& \fr{\pi^2}{b_1} \int_{\om > z} \fr{d^3 \bk}{(2\pi)^3} \fr{1}{\om^3} 
                \left\{ 1 + i\om \left(\eta-\eta^\pp \right) \right\} 
                  e^{ -i\om(\eta -\eta^\pp -i\eps)+ i \bk \cdot (\bx -\bx^\pp) } 
            \nonumber \\  
        &=& -\fr{1}{4b_1} \log \left\{ \left[ -(\eta-\eta^\pp-i\eps)^2 + (\bx -\bx^\pp)^2 \right]  
                                               z^2 e^{2\gm-2} \right\} 
              \nonumber \\
        && -\fr{1}{4b_1}\fr{i\eps}{|\bx-\bx^\pp|}
            \log \fr{\eta-\eta^\pp-i\eps-|\bx-\bx^\pp|}{\eta-\eta^\pp-i\eps+|\bx-\bx^\pp|} .
               \label{phi-phi correlation function}
\end{eqnarray}
Here, $\eps$ is the cutoff parameter to regularize UV divergences. Since the Riegert field is dimensionless, we also introduce an infinitesimal mass scale $z$ to handle IR divergences.\footnote{ 
It corresponds to add a fictitious mass term to the action, which is not gauge invariant so that the $z$-dependence should cancel out when we consider physical quantities. Here, note that the Einstein and the cosmological constant terms cannot be considered to be the mass term due to the presence of the exponential factor of the Riegert field in our formalism. 
}  
The integration is carried out under the condition $z \ll 0$, while the UV cutoff $\eps$ is finite (see Appendix B). Until all calculations are finished, $\eps$ does not take zero.

The singular parts for other field variables are also obtained from the field operators (\ref{expansion of chi and momentum variables}), or by differentiating the correlation function (\ref{phi-phi correlation function}) according to the definitions of field variables. They are summarized in Appendix C for the equal-time cases used below. Using these expressions, the canonical commutation relation (\ref{canonical commutation relation}), for instance, can be expressed as 
\begin{eqnarray}
    \left[ \phi(\eta, \bx), \P_\phi(\eta, \bx^\pp) \right] 
    &=& \lang \phi(\eta, \bx) \P_\phi(\eta, \bx^\pp) \rang - \lang \phi(\eta, \bx) \P_\phi(\eta, \bx^\pp) \rang^\dag
           \nonumber \\
    &=& i \fr{1}{\pi^2} \fr{\eps}{[(\bx -\bx^\pp)^2 +\eps^2]^2} ,
\end{eqnarray}
where the last term is nothing but the $\dl$-function regularized to be
\begin{eqnarray}
   \dl_3(\bx) = \int \fr{d^3{\bf k}}{(2\pi)^3} e^{i\bk \cdot \bx-\eps \om} = \fr{1}{\pi^2} \fr{\eps}{(\bx^2 +\eps^2)^2} .
         \label{regularized delta function}
\end{eqnarray}

Applying the OPE technique to the composite operators, we obtain the following formula:
\begin{eqnarray}
   && [ \l: AB(x) \r:, \l: C^n(y) \r: ] 
          \nonumber \\
   && = n [ A(x), C(y) ] \l: B(x)C^{n-1}(y) \r: + n [ B(x), C(y)] \l: A(x)C^{n-1}(y) \r: 
                 \nonumber \\
   && \quad
       + n(n-1) \left\{ \lang A(x)C(y) \rang \lang B(x)C(y) \rang 
                          - h.c. \right\} \l: C^{n-2}(y) \r: .
\end{eqnarray}
The last term is the quantum correction and $h.c.$ denotes the hermitian conjugate of the first term in the bracket. The quantum correction term thus vanishes if it is real.

Now, we compute the transformation law of the composite operator $\l: \phi^n \r:$. We first calculate the equal-time commutation relations between this operator and the local operators that appear in the generators. For the Hamiltonian density $\calA$, we obtain
\begin{eqnarray}
    \left[ \calA(\bx), \l: \phi^n(\by) \r: \right]  &=& -i n \dl_3 (\bx-\by ) \l: \chi \phi^{n-1} (\by) \r: 
            \nonumber \\
                  &=& -i \dl_3 (\bx -\by) \pd_\eta \l: \phi^n(\by) \r: .
         \label{comutator between A and phi^n}
\end{eqnarray}
The $\dl$-function comes from the commutator between $\phi$ and $\P_\phi$ and quantum corrections vanish due to $\langle \chi(\bx) \phi(\by) \rangle=0$ and so on. 
Here and below, when we discuss equal-time commutation relations we do not write the time-coordinate dependence in field variables for simplicity.

In the equal-time commutator with $\calB_j$, the quantum correction survives such that
\begin{eqnarray}
      \left[ \calB_j(\bx), \l: \phi^n(\by) \r: \right] 
        &=& -i \dl_3 (\bx -\by ) \pd_j \l: \phi^n (\by) \r: 
              \nonumber \\
     &&  + i \fr{1}{2b_1} n(n-1) e_j(\bx-\by) \l: \phi^{n-2}(\bx) \r: ,
             \label{commutator between B_j and phi^n}
\end{eqnarray}
where the function $e_j(\bx)$ denotes the quantum correction, defined by
\begin{equation}
     e_j(\bx) = \fr{1}{\pi^2} \fr{\eps x_j [1-h(\bx)]}{\bx^2(\bx^2 + \eps^2)^2} , \quad
     h(\bx) = \fr{i\eps}{2|\bx|}\log \fr{i\eps + |\bx|}{i\eps -|\bx|} ,
          \label{function e_j and h}
\end{equation}
where $h^\dag(\bx)=h(\bx)$ and $\lim_{\bx \to 0} h(\bx) = 1$.

Since the generator of conformal symmetry is time-independent, we can compute commutators between the generators and field operators using equal-time commutators. From (\ref{comutator between A and phi^n}) and (\ref{commutator between B_j and phi^n}) and also $[\l: \P_\chi \pd_j \phi(\bx) \r:, \l: \phi^n(\by) \r:] =0$, we find the commutators for translations and Lorentz transformations to be
\begin{eqnarray}
   i \left[ P_\mu, \l: \phi^n(x) \r: \right] &=& \pd_\mu \l: \phi^n(x) \r: ,
             \nonumber \\
   i \left[ M_{\mu\nu}, \l: \phi^n(x) \r: \right] &=& \left( x_\mu \pd_\nu - x_\nu \pd_\mu \right) \l: \phi^n(x) \r: .
          \label{transformation law of phi^n I}
\end{eqnarray}
Here, the quantum correction term cancels out. It is shown by using the following integral formulae: $\int \d3x  e_j(\bx)=0$ due to the odd property under $\bx \to -\bx$ and
\begin{eqnarray}
    \int \d3x x_i e_j(\bx) 
     = \fr{1}{3} \dl_{ij} \int^\infty_0 4\pi x^2 dx \fr{1}{\pi^2} \fr{\eps [1-h(x)]}{(x^2 + \eps^2)^2} 
        = \fr{1}{6} \dl_{ij} .
      \label{integral of x_i e_j}
\end{eqnarray}
The antisymmetric property of the Lorentz generator is also used.

In the same way, we find that the commutators for dilatations and special conformal transformations are given by
\begin{eqnarray}
   i \left[ D, \l: \phi^n(x) \r: \right] &=& 
      x^\mu \pd_\mu \l: \phi^n(x) \r: + n \l: \phi^{n-1}(x) \r: -\fr{1}{4b_1}n(n-1) \l: \phi^{n-2}(x) \r: ,
             \nonumber \\
   i \left[ K_\mu, \l: \phi^n(x) \r: \right] 
       &=& \left( x^2 \pd_\mu - 2x_\mu x^\nu \pd_\nu \right) \l: \phi^n(x) \r: 
             \nonumber \\
       && - 2x_\mu \left( n \l: \phi^{n-1}(x) \r: - \fr{1}{4b_1}n(n-1) \l: \phi^{n-2}(x) \r: \right) .
          \label{transformation law of phi^n II}
\end{eqnarray}
Here, the $\l: \phi^{n-1} \r:$ terms come from the commutation relations with the linear $\P_\phi$ terms in these generators. The last terms with $1/b_1$ are the quantum corrections. For the commutator with $K_j$, we use the formula, 
\begin{equation}
    \int \d3x \left\{ \bx^2 e_j(\bx-\by) -2 x_j x^k e_k(\bx-\by) \right\} = - y_j
\end{equation} 
derived from (\ref{integral of x_i e_j}).

The transformation of $n=1$ is nothing but the gauge transformation (\ref{conformal transformation}) such that $i[Q_\zeta, \phi] = \dl_\zeta \phi$. In this case, the quantum correction term vanishes.

The most simple conformal field is given by the exponential function of the Riegert field defined as
\begin{equation}
      V_\a (x) = \l: e^{\a\phi(x)} \r: = \sum_{n=0}^\infty \fr{\a^n}{n!} \l: \phi^n(x) \r: .
           \label{conformal field V_a}
\end{equation}
From the transformation properties of (\ref{transformation law of phi^n I}) and (\ref{transformation law of phi^n II}), we obtain the following conformal transformations:
\begin{eqnarray}
      i\left[ P_\mu, V_\a(x) \right] &=& \pd_\mu V_\a(x) ,
             \nonumber \\
   i \left[ M_{\mu\nu}, V_\a(x) \right] &=& \left( x_\mu \pd_\nu - x_\nu \pd_\mu \right) V_\a (x) ,
             \nonumber \\ 
   i\left[ D, V_\a(x) \right] &=& \left( x^\mu \pd_\mu + h_\a \right) V_\a(x) ,
             \nonumber \\
   i \left[ K_\mu, V_\a (x) \right] 
       &=& \left( x^2 \pd_\mu - 2x_\mu x^\nu \pd_\nu -2 x_\mu h_\a \right) V_\a (x) ,
            \label{conformal transformation of V_a}
\end{eqnarray}
where $h_\a$ is the conformal dimension calculated to be
\begin{equation}
     h_\a = \a -\fr{\a^2}{4b_1} .
\end{equation}
The second term proportional to $1/b_1$ is the quantum correction, which coincides with the result computed on the $R \times S^3$ background \cite{amm97, hamada11}.

\subsection{Physical Fields}
\noindent

Conformal fields themselves are not diffeomorphism invariant, namely, they are not physical fields. Physical fields are defined by field operators that commute with all generators of conformal symmetry as
\begin{eqnarray}
    \left[ Q_\zeta, \int d^4 x {\cal O}(x) \right] = 0 .
         \label{physical field condition}
\end{eqnarray}

The most simple example of the local operator ${\cal O}$ is given by $V_\a$ with conformal dimension $h_\a=4$ because in this case the conformal transformation (\ref{conformal transformation of V_a}) becomes 
\begin{eqnarray}
    i \left[ Q_\zeta , V_\a(x) \right] = \pd_\mu \left\{ \zeta^\mu V_\a(x) \right\} 
        \label{physical transformation law}
\end{eqnarray}
for all conformal Killing vectors $\zeta^\mu$ (\ref{conformal Killing vector}).

Since the constant $\a$ obtained as the solution of $h_\a =4$ is real due to $b_1 >4$ independently of matter field contents (see footnote 6), the operator $V_\a$ becomes real. This physical field corresponds to the cosmological constant term.

In general, ${\cal O}$ is given by scalar fields with conformal dimension $4$ composed of Riegert fields as well as traceless tensor fields, but fields that tensor indices retain are excluded from physical fields because  even though they have conformal dimension $4$, their transformations cannot be written in the form (\ref{physical transformation law}) due to the presence of spin terms in $M_{\mu\nu}$ and $K_\mu$.

The positivity of the two-point function of $V_\a$ is trivial because it is a real operator. Since diffeomorphism invariant fields will be real, the Wightman positivity condition for these fields seems to be satisfied. Of course, in this argument, it is crucial that both Riegert-Wess-Zumino and Weyl actions have a correct sign such that they are bounded from below (if we do the usual analytic continuation to Euclidean signature).

\section{Conformal Algebra}
\setcounter{equation}{0}
\noindent

Now, we can show that the generators presented above indeed form conformal algebra at the quantum level.

First, we present the useful OPE formula, 
\begin{eqnarray}
   \left[ \l: AB(x) \r:, \l: CD(y) \r: \right] 
    &=& \left[ A(x), C(y) \right] \l: B(x)D(y) \r: + \left[ A(x), D(y) \right] \l: B(x)C(y) \r:
             \nonumber \\
    && + \left[ B(x), C(y) \right] \l: A(x)D(y) \r: + \left[ B(x), D(y) \right] \l: A(x)C(y) \r:
             \nonumber \\
    &&   + {\rm q.c.}(x-y) .
\end{eqnarray}
The last term is the quantum correction given by
\begin{eqnarray}
     {\rm q.c.}(x-y) = \lang A(x)C(y) \rang \lang B(x)D(y) \rang 
                 + \lang A(x)D(y) \rang \lang B(x)C(y) \rang 
                 - h.c. ,
\end{eqnarray}
where $h.c.$ denotes the hermitian conjugate of the first two terms.

Using this formula, the equal-time commutation relations among the operators $\calA$ and $\calB_j$ are calculated as
\begin{eqnarray}
    \left[ \calA(\bx), \calA(\by) \right]
     &=& i \lap3_x \dl_3 (\bx-\by) \Bigl( \l: \P_\chi(\bx) \chi(\by) \r: - \l: \chi(\bx)\P_\chi(\by) \r: \Bigr)
            \nonumber \\
     && + i \fr{b_1}{4\pi^2} \lap3_x \dl_3 (\bx-\by) 
            \Bigl( \l: \lap3 \phi(\bx) \chi(\by) \r: - \l: \chi(\bx) \lap3 \phi(\by) \r: \Bigr) ,
          \nonumber \\
    \left[ \calB_j(\bx), \calB_k(\by) \right]
     &=& i \pd_j^x \dl_3 (\bx-\by) \Bigl( \l: \P_\chi(\bx) \pd_k \chi(\by) \r: 
                             + \l: \P_\phi(\bx) \pd_k \phi(\by) \r: \Bigr)
           \nonumber \\
     &&  + i \pd_k^x \dl_3 (\bx-\by) \Bigl( \l: \pd_j \chi(\bx) \P_\chi(\by) \r:
                             + \l: \pd_j \phi(\bx) \P_\phi(\by) \r: \Bigr) 
           \label{AA and BB commutators}
\end{eqnarray}
and 
\begin{eqnarray}
     \left[ \calA(\bx), \calB_j(\by) \right]
     &=& i \pd_j^x \dl_3(\bx-\by)\left( - \fr{4\pi^2}{b_1} \l: \P_\chi(\bx)\P_\chi(\by) \r: 
                                        + \l: \chi(\bx)\P_\phi(\by) \r: \right)
               \nonumber \\
     &&  + i \dl_3(\bx-\by) \left( \l: \P_\phi \pd_j \chi(\by) \r: 
                  + \fr{b_1}{4\pi^2} \l: \lap3 \chi \pd_j \chi(\by) \r: \right)
             \nonumber \\
     &&  + i \fr{b_1}{4\pi^2} \lap3_x \dl_3(\bx-\by) \Bigl( \l: \chi(\bx)\pd_j \chi(\by) \r:
                       + \l: \lap3 \phi(\bx) \pd_j \phi(\by) \r: \Bigr)
             \nonumber \\
     &&    - i \fr{4}{\pi^2} f_j(\bx-\by) ,
            \label{AB commutator}
\end{eqnarray}
where the function $f_j$ denotes the quantum correction, defined by
\begin{equation}
    f_j (\bx) = \fr{1}{\pi^2} \fr{\eps x_j (\bx^2 -\eps^2)}{(\bx^2 +\eps^2)^6} .
          \label{function f_j}              
\end{equation}

Here, we summarize the properties of this function. It can be written in the derivative of scalar function as $f_j(\bx) = \pd_j f(\bx)$, where 
\begin{eqnarray}
            f(\bx) = -\fr{1}{40\pi^2}\fr{\eps(5\bx^2 -3\eps^2)}{(\bx^2 +\eps^2)^5} .
                  \label{scalar function f}
\end{eqnarray}
The integrals of these functions satisfy
\begin{eqnarray}
 \int d^3 \bx f_j(\bx) =0,  \quad \int d^3 \bx f(\bx) =0, \quad \int d^3 \bx x^j f(\bx) =0 ,
        \label{property of integrals of f}
\end{eqnarray}
while $\int d^3 \bx \bx^2 f(\bx)=-1/160\eps^2$ diverges at the limit $\eps \to 0$.\footnote{ 
The function $f$ can be written in terms of the delta function as
$$ f(\bx)= -\fr{1}{320} \lap3 \left( \fr{1}{\bx^2} \dl_3 (\bx) \right). $$
To derive this, we use the expression of the regularized delta function (\ref{regularized delta function}) and its variant form $\pi^2 \dl_3(\bx) = 4\eps^3/(\bx^2+\eps^2)^3$.  
} 

\subsection{Poincar\'{e} Algebra}
\noindent

Let us first show that the Poincar\'{e} algebra is closed. Using the equal-time commutation relations computed above, we obtain the following commutators:
\begin{eqnarray}
   \left[ P_0, \calA(x) \right] &=& - i \pd_\eta \calA(x) ,  \quad
   \left[ P_0, \calB_j(x) \right] = -i \pd_\eta \calB_j(x),
            \nonumber \\
   \left[ P_j, \calA(x) \right] &=& -i \pd_j \calA(x) ,  \quad
   \left[ P_j, \calB_k(x) \right] = -i \pd_j \calB_k (x) .   
\end{eqnarray}
Here, the time-derivatives of $\calA$ and $\calB_j$ are defined from the equations of motion by
\begin{eqnarray}
   \pd_\eta \calA &=&  \l: \P_\chi \lap3 \chi \r: - \l: \chi \lap3 \P_\chi \r: 
        +\fr{b_1}{4\pi^2} \left( \l: \lap3 \chi \lap3 \phi \r: - \l: \chi \dlap3 \phi \r: \right) ,
              \nonumber \\
   \pd_\eta \calB_j &=& -\fr{4\pi^2}{b_1} \l: \P_\chi \pd_j \P_\chi \r: 
        - \fr{b_1}{4\pi^2} \left( 2 \l: \lap3 \chi \pd_j \chi \r: + \l: \dlap3 \phi \pd_j \phi \r: \right) 
\end{eqnarray}
and the quantum correction term disappears due to the property of $f_j$ (\ref{property of integrals of f}).

In order to obtain the Lorentz algebra, we further calculate the following equal-time commutation relations:
\begin{eqnarray}
     \left[ \calA(\bx), \l: \P_\chi \pd_j \phi(\by) \r: \right]
     &=&   i \dl_3(\bx-\by) \left( \l: \P_\phi \pd_j \phi(\by) \r: 
                  + \fr{b_1}{4\pi^2} \l: \lap3 \chi \pd_j \phi(\by) \r: \right) 
             \nonumber \\
     && + i \pd_j^x \dl_3(\bx-\by) \l: \chi(\bx)\P_\chi(\by) \r: 
             \nonumber \\
     && + i \fr{b_1}{4\pi^2} \lap3_x \dl_3(\bx-\by) \l: \chi(\bx)\pd_j \phi(\by) \r: ,
               \nonumber \\
     \left[ \calB_j(\bx), \l: \P_\chi \pd_k \phi(\by) \r: \right]
     &=& i \pd_j^x \dl_3(\bx-\by) \l: \P_\chi(\bx) \pd_k \phi(\by) \r: 
              \nonumber \\
      &&    + i \pd_k^x \dl_3(\bx-\by) \l: \pd_j \phi(\bx) \P_\chi(\by) \r: .
\end{eqnarray}
Using (\ref{AA and BB commutators}), (\ref{AB commutator}) and these equations, we obtain the following algebra for the generators of Lorentz transformations:
\begin{eqnarray}
    \left[ M_{0j}, \calA(x) \right] &=&
      i \eta \pd_j \calA(x)   + i x_j \pd_\eta \calA(x)
              \nonumber \\
      && + i \biggl\{ \l: \P_\phi \pd_j \phi (x) \r: + 2 \l: \P_\chi \pd_j \chi(x) \r: - \l: \chi \pd_j \P_\chi(x) \r: 
              \nonumber \\
      &&  + \fr{b_1}{4\pi^2} \Bigl( \l: \lap3 \chi \pd_j \phi(x) \r: + 2 \l: \pd_j \chi \lap3 \phi(x) \r:
                                   - \l: \chi \pd_j \lap3 \phi(x) \r: \Bigr)  \biggr\},
               \nonumber \\
    \left[ M_{0j}, \calB_k(x) \right] &=& 
      i \eta \pd_j \calB_k (x) + i x_j \pd_\eta \calB_k(x)
               \nonumber \\
      && + i \biggl\{ \dl_{jk} \left( -\fr{4\pi^2}{b_1} \l: \P_\chi^2(x) \r: + \l: \P_\phi \chi(x) \r: \right)
              + \l: \P_\chi \pd_j \pd_k \phi(x) \r: 
               \nonumber \\
      &&  + \l: \pd_j \P_\chi \pd_k \phi(x) \r: 
           - \fr{b_1}{2\pi^2} \Bigl( \l: \pd_j \chi \pd_k \chi(x) \r: 
                         + \l: \pd_j \lap3 \phi \pd_k \phi(x) \r:  \Bigr)  \biggr\} ,
               \nonumber \\
    \left[ M_{0j}, \l: \P_\chi \pd_k \phi (x) \r: \right] &=& 
      i \eta \left( \l: \pd_j \P_\chi \pd_k \phi(x) \r: + \l: \P_\chi \pd_j \pd_k \phi(x) \r: \right)
              \nonumber \\
      && + i x_j \left( \l: \P_\chi \pd_k \chi(x) \r: - \l: \P_\phi \pd_k \phi(x) \r:  
                      - \fr{b_1}{2\pi^2} \l: \lap3 \chi \pd_k \phi(x) \r: \right)
               \nonumber \\
      && + i \dl_{jk} \l: \P_\chi \chi(x) \r: -i \fr{b_1}{2\pi^2} \l: \pd_j \chi \pd_k \phi(x) \r: 
\end{eqnarray}
and
\begin{eqnarray}
    \left[ M_{ij}, \calA(x) \right] &=& 
          -i x_i \pd_j \calA(x) + i x_j \pd_i \calA(x) ,
               \nonumber \\
    \left[ M_{ij}, \calB_k(x) \right] &=& 
          - i x_i \pd_j \calB_k(x) + i x_j \pd_i \calB_k(x) 
          - i \dl_{ik} \calB_j(x) + i \dl_{jk} \calB_i(x) ,
               \nonumber \\
    \left[ M_{ij}, \l: \P_\chi \pd_k \phi (x) \r: \right] &=& 
          - i x_i \Bigl( \l: \pd_j \P_\chi \pd_k \phi(x) \r: + \l: \P_\chi \pd_j \pd_k \phi(x) \r: \Bigr)
          - i \dl_{ik} \l: \P_\chi \pd_j \phi(x) \r:
               \nonumber \\
       && + i x_j \Bigl( \l: \pd_i \P_\chi \pd_k \phi(x) \r: + \l: \P_\chi \pd_i \pd_k \phi(x) \r: \Bigr)
          + i \dl_{jk} \l: \P_\chi \pd_i \phi(x) \r: .
              \nonumber \\
\end{eqnarray}
Here, all quantum correction terms vanish due to (\ref{property of integrals of f}).

Using these algebra, we can show that the Poincar\'{e} algebra
\begin{eqnarray}
    \left[ P_\mu, P_\nu \right] &=& 0 ,  \qquad
    \left[ M_{\mu\nu}, P_\lam \right] = -i \left( \eta_{\mu\lam} P_\nu - \eta_{\nu\lam} P_\mu \right),
          \nonumber \\
    \left[ M_{\mu\nu}, M_{\lam\s} \right] &=& 
        -i \left( \eta_{\mu\lam}M_{\nu\s} + \eta_{\nu\s}M_{\mu\lam} 
                  - \eta_{\mu\s}M_{\nu\lam} - \eta_{\nu\lam}M_{\mu\s} \right) 
          \label{Poincare algebra}
\end{eqnarray}
is satisfied at the quantum level.

\subsection{Conformal Algebra with $D$ and $K_\mu$}
\noindent

Next, we calculate the remaining algebra including the generators of dilatations and special conformal transformations, $D$ and $K_\mu$. In order to calculate it, we further need the following equal-time commutation relations:
\begin{eqnarray}
     \left[ \calA(\bx), \l: \P_\chi \chi(\by) \r: \right]
     &=&  i \dl_3(\bx-\by) \left( \fr{4\pi^2}{b_1} \l: \P^2_\chi(\by) \r: 
          + \l: \P_\phi \chi(\by) \r: + \fr{b_1}{4\pi^2} \l: \chi \lap3 \chi(\by) \r: \right) 
             \nonumber \\
     && + i \fr{b_1}{4\pi^2} \lap3_x \dl_3(\bx-\by) \l: \chi(\bx) \chi(\by) \r: 
        + i \fr{10}{\pi^2} f(\bx-\by),
               \nonumber \\
     \left[ \calA(\bx), \l: \chi^2(\by) \r: \right]
     &=&  i \fr{8\pi^2}{b_1} \dl_3(\bx-\by) \l: \P_\chi \chi(\by) \r: ,
             \nonumber \\
     \left[ \calA(\bx), \l: \pd_k \phi \pd^k \phi(\by) \r: \right]
     &=& 2i \pd_k^x \dl_3(\bx-\by) \l: \chi(\bx) \pd^k \phi(\by) \r: ,
             \nonumber \\
     \left[ \calA(\bx), \l: \chi \pd_j \phi(\by) \r: \right]
     &=& i \fr{4\pi^2}{b_1} \dl_3(\bx-\by) \l: \P_\chi \pd_j \phi(\by) \r:
         + i \pd_j^x \dl_3(\bx-\bx^\pp) \l: \chi(\bx) \chi(\by) \r:
             \nonumber \\
      && + i \fr{1}{b_1} g_j(\bx-\by)  
\end{eqnarray}
and
\begin{eqnarray}
     \left[ \calB_j(\bx), \l: \P_\chi \chi(\by) \r: \right]
     &=& - i \dl_3(\bx-\by) \l: \P_\chi \pd_j \chi(\by) \r: 
          + i \pd_j^x \dl_3(\bx-\by) \l: \P_\chi(\bx) \chi(\by) \r: ,
              \nonumber \\
     \left[ \calB_j(\bx), \l: \chi^2(\by) \r: \right]
     &=& -2i \dl_3(\bx-\by) \l: \chi \pd_j \chi(\by) \r: 
         - i \fr{2}{b_1} g_j(\bx-\by) ,
              \nonumber \\
     \left[ \calB_j(\bx), \l: \pd_k \phi \pd^k \phi(\by) \r: \right]
     &=& 2 i \pd_k^x \dl_3(\bx-\by) \l: \pd_j \phi(\bx) \pd^k \phi(\by) \r: 
              \nonumber \\
      &&   -i \fr{4}{b_1} \left\{ g_j(\bx-\by) - {\tilde g}_j(\bx-\by) \right\}  ,
               \nonumber \\
     \left[ \calB_j(\bx), \l: \chi \pd_k \phi(\by) \r: \right]
     &=& - i \dl_3(\bx-\by) \l: \pd_j \chi \pd_k \phi(\by) \r: 
          +i \pd_k^x \dl_3(\bx-\by) \l: \pd_j \phi(\bx) \chi(\by) \r: . 
              \nonumber \\
\end{eqnarray}
Here, the function $f$ is defined by (\ref{scalar function f}) and we also introduce other functions defined by
\begin{eqnarray}
    g_j (\bx) = \fr{1}{\pi^2} \fr{\eps x_j}{(\bx^2 +\eps^2)^4} 
      \label{function g_j}
\end{eqnarray}
and 
\begin{eqnarray}
    {\tilde g}_j(\bx) = \fr{2}{\pi^2} \fr{\eps x_j[1 -h(\bx)]}{\bx^2 (\bx^2 +\eps^2)^3} ,
          \label{function tilde g_j}
\end{eqnarray}
where $h$ is defined in (\ref{function e_j and h}). The function $g_j$ can be written in the derivative of scalar function as $g_j(\bx)=\pd_j g(\bx)$, where 
\begin{eqnarray}
    g(\bx) = -\fr{1}{6\pi^2} \fr{\eps}{(\bx^2 +\eps^2)^3} . 
          \label{scalar function g}
\end{eqnarray}
This function is related to the scalar function (\ref{scalar function f}) as $\lap3 g(\bx)=40f(\bx)$.

Since $g_j$ and ${\tilde g}_j$ are odd under the change $\bx \to -\bx$, the integrals of them vanish such that $\int \d3x g_j(\bx)=\int \d3x {\tilde g}_j(\bx)=0$, while the integral of $g$ is given by $\int \d3x g(\bx) =-1/24\eps^2$, which diverges at the limit of $\eps \to 0$. Thus, the following integral diverges at the limit:
\begin{equation}
   \int \d3x x_i g_j(\bx) = - \dl_{ij} \int \d3x g(\bx) 
         = \dl_{ij} \fr{1}{24\eps^2} .
\end{equation}
We also obtain the divergent integral,
\begin{equation}
   \int \d3x x_i {\tilde g}_j(\bx) = \fr{1}{3}\dl_{ij} \int^\infty_0 4\pi x^2 dx \fr{2\eps [1-h(x)]}{\pi^2 (x^2+\eps^2)^3} 
      =\dl_{ij} \fr{1}{24\eps^2} .
\end{equation}
The difference of these two integrals, however, vanishes exactly such that we find
\begin{eqnarray}
    \int \d3x x_i \left\{ g_j(\bx) -{\tilde g}_j(\bx) \right\} = 0 .
\end{eqnarray}
This formula is used when we show that quantum corrections disappear.

Furthermore, we need the following equal-time commutation relations:
\begin{eqnarray}
    \left[ \l: \P_\chi \chi(\bx) \r:, \l: \P_\chi \pd_j \phi(\by) \r: \right]
      &=&  i \dl_3 (\bx-\by) \l: \P_\chi \pd_j \phi(\by) \r: - i \fr{1}{4\pi^2}g_j(\bx-\by), 
             \nonumber \\
    \left[ \l: \P_\chi \pd_j \phi(\bx) \r:, \l: \chi^2(\by) \r: \right]
      &=&  - 2i \dl_3 (\bx-\by) \l: \chi \pd_j \phi(\by) \r: , 
             \nonumber \\
    \left[ \l: \P_\chi \pd_j \phi(\bx) \r:, \l: \chi \pd_k \phi(\by) \r: \right]
      &=&  - i \dl_3 (\bx-\by) \l: \pd_j \phi \pd_k \phi(\by) \r: 
            - i \fr{1}{2b_1} g_{jk}(\bx-\by) , 
             \nonumber \\
    \left[ \l: \P_\chi \chi(\bx) \r:, \l: \chi^2(\by) \r: \right]
      &=&  -2i \dl_3 (\bx-\by) \l: \chi^2(\by) \r: - i \fr{6}{b_1}g(\bx-\by), 
             \nonumber \\
    \left[ \l: \P_\chi \chi(\bx) \r:, \l: \chi \pd_j \phi(\by) \r: \right]
      &=&  - i \dl_3 (\bx-\by) \l: \chi \pd_j \phi(\by) \r: , 
\end{eqnarray}
where the function $g_{ij}$ in the third equation is defined by
\begin{equation}
    g_{ij}(\bx) = \fr{1}{\pi^2} \left\{ \fr{x_i x_j}{\bx^2}\fr{\eps}{(\bx^2 +\eps^2)^3}
                    + \left( \dl_{ij} - \fr{3 x_i x_j}{\bx^2} \right) 
                          \fr{\eps [1-h(\bx)]}{\bx^2 (\bx^2 + \eps^2)^2} \right\} .
\end{equation}
The integral of this function is given by
\begin{eqnarray}
    \int \d3x g_{ij}(\bx) = -2\dl_{ij} \int \d3x g(\bx) ,
\end{eqnarray}
which diverges at the limit of $\eps \to 0$.

From the commutators obtained above, we find that the commutators between the generator $D$ and various local operators are given by
\begin{eqnarray}
    \left[ D, \calA(x) \right] &=& 
      - i \left( \eta \pd_\eta + x^k \pd_k +4 \right) \calA(x) ,
            \nonumber \\
    \left[ D, \calB_j(x) \right] &=& 
       - i \left( \eta \pd_\eta + x^k \pd_k +4 \right) \calB_j(x)
\end{eqnarray}
and
\begin{eqnarray}
    \left[ D, \l: \P_\chi \pd_j \phi(x) \r: \right] &=& 
        - i \left( \eta \pd_\eta + x^k \pd_k + 3 \right) \l: \P_\chi \pd_j \phi(x) \r: ,
            \nonumber \\
    \left[ D, \l: \P_\chi \chi(x) \r: \right] &=& 
        - i \left( \eta \pd_\eta + x^k \pd_k + 3 \right) \l: \P_\chi \chi(x) \r: ,
            \nonumber \\
    \left[ D, \l: \chi^2(x) \r: \right] &=& 
       - i \left( \eta \pd_\eta + x^k \pd_k +2 \right) \l: \chi^2(x) \r: ,
           \nonumber \\
    \left[ D, \l: \pd_k \phi \pd^k \phi(x) \r: \right] &=& 
       - i \left( \eta \pd_\eta + x^k \pd_k +2 \right) \l: \pd_k \phi \pd^k \phi(x) \r: ,
            \nonumber \\   
    \left[ D, \l: \chi \pd_j \phi(x) \r: \right] &=& 
       - i \left( \eta \pd_\eta + x^k \pd_k +2 \right) \l: \chi \pd_j \phi(x) \r: .
\end{eqnarray}
Here, all divergent terms cancel out and quantum corrections disappear, which can be shown by using the integral formulae for various functions presented above.

In the same way, we can calculate the commutation relations between the generator $K_\mu$ and various local operators, which are summarized in Appendix D. From these commutators, we find that the following algebra is satisfied without quantum correction terms:
\begin{eqnarray}
    \left[ D, P_\mu \right] &=& -i P_\mu , \quad \left[ D, M_{\mu\nu} \right] = 0, 
             \quad \left[ D, K_\mu \right] = i K_\mu , 
          \nonumber \\
    \left[ M_{\mu\nu}, K_\lam \right] &=& -i \left( \eta_{\mu\lam} K_\nu - \eta_{\nu\lam} K_\mu \right),
             \quad \left[ K_\mu, K_\nu \right] = 0 ,
          \nonumber \\
    \left[ K_\mu, P_\nu \right] &=& 2i \left( \eta_{\mu\nu}D + M_{\mu\nu} \right) .
\end{eqnarray}
Combined with the Poincar\'{e} algebra (\ref{Poincare algebra}), we obtain the conformal algebra.

\section{Traceless Tensor Fields}
\setcounter{equation}{0}
\noindent

In the UV limit of $t \to 0$, the Weyl action $-(1/t^2)\int d^4 x \sq{-g}C^2_{\mu\nu\lam\s}$ reduces to the quadratic form, 
\begin{eqnarray}
     I = \int d^4 x \left\{ -\half \pd^2 h^{\mu\nu} \pd^2 h_{\mu\nu}  
                            + \pd^\mu \chi^\nu \pd_\mu \chi_\nu  
                            - \fr{1}{3} \pd_\mu \chi^\mu \pd_\nu \chi^\nu  \right\} ,
\end{eqnarray}
where $\chi_\mu = \pd^\lam h_{\lam \mu}$. This action is invariant under the gauge transformations $\dl_\kappa h_{\mu\nu}$ (\ref{gauge transformation}) and $\dl_\zeta h_{\mu\nu}$ (\ref{conformal transformation}).

In this section, after the traceless tensor field is quantized, we derive the generator of conformal symmetry and then discuss the transformation law of the field.

\subsection{Canonical Quantization}
\noindent

In order to quantize the traceless tensor field, we have to fix the gauge symmetry $\dl_\kappa h_{\mu\nu}$ (\ref{gauge transformation}). Using four gauge degrees of freedom $\kappa^\mu$, we can take the Coulomb gauge defined by $\pd_i h^i_{~\mu}=0$. The $(00)$ component $h_{00}(=h^i_{~i})$ then becomes nondynamical. Therefore, using the residual gauge symmetry preserving the Coulomb gauge conditions, we further remove the $h_{00}$ field. This is the radiation gauge summarized by the conditions,  
\begin{eqnarray}
    h_{00}=0, \quad  h_{0j} = \h_j,  \quad h_{ij} = \h_{ij} ,   
         \label{radiation gauge} 
\end{eqnarray}
where $\h_j$ and $\h_{ij}$ are the transverse vector mode and the transverse traceless tensor mode, respectively, satisfying the conditions $\pd^i \h_i =0$ and $\pd^i \h_{ij} = \h^i_{~i} = 0$.

In the radiation gauge, the gauge degrees of freedom $\kappa^\mu$ are completely fixed and thus only the gauge symmetry $\dl_\zeta h_{\mu\nu}$ (\ref{conformal transformation}) survives as the residual gauge symmetry.

The transverse traceless tensor mode is quantized according to the Dirac's procedure by introducing new variable,
\begin{equation}
    \u_{ij} = \pd_\eta \h_{ij} ,
\end{equation}
while the transverse vector mode is treated as it is because it is the second order with respect to the time derivative. 
The Weyl action is then written in the form,
\begin{eqnarray}
     I &=& \int d^4 x \left\{ - \half \h^{ij} \left( \pd_\eta^4 - 2\lap3 \pd_\eta^2 +\dlap3 \right) \h_{ij}
                                + \h^j \lap3 \left( -\pd_\eta^2 + \lap3 \right) \h_j  \right\} 
           \nonumber \\
       &=& \int d^4 x \biggl\{ 
              - \half \pd_\eta \u^{ij} \pd_\eta \u_{ij} - \u^{ij} \lap3 \u_{ij} 
              - \half \lap3 \h^{ij} \lap3 \h_{ij}  + \lam^{ij} \left( \pd_\eta \h_{ij} - \u_{ij} \right)
           \nonumber \\
      && \qquad\qquad
             + \pd_\eta \h^j \lap3 \pd_\eta \h_j  + \lap3 \h^j \lap3 \h_j
              \biggr\} ,
\end{eqnarray}
where $\lam^{ij}$ is the Lagrange multiplier.

After removed the variable $\lam^{ij}$ by solving the constraint equation, we obtain the phase space spanned by six canonical variables $\u_{ij}$, $\h_{ij}$, $\h_j$, and their conjugate momenta,
\begin{eqnarray}
     \P_\u^{ij} &=& - \pd_\eta \u^{ij}, 
                \nonumber \\
     \P_\h^{ij} &=& - \pd_\eta \P_\u^{ij} - 2 \lap3 \u^{ij}, 
                \nonumber \\
     \P^j &=& 2 \lap3 \pd_\eta \h^j .
\end{eqnarray}
The canonical commutation relations are set as
\begin{eqnarray}
     \left[ \h^{ij}(\eta, \bx), \P_\h^{kl}(\eta, \by) \right] 
       &=& \left[ \u^{ij}(\eta, \bx), \P_\u^{kl}(\eta, \by) \right] = i \dl^{ij,kl}_3 (\bx -\by),
           \nonumber \\
     \left[ \h^i (\eta, \bx), \P^j(\eta, \by) \right] 
       &=& i \dl^{ij}_3 (\bx -\by) ,  
          \label{equal-time commutation relation for tensor field}    
\end{eqnarray}
where the delta functions are defined by $\dl_3^{ij}(\bx) = \Delta^{ij} \dl_3(\bx)$ and $\dl_3^{ij,kl}(\bx) = \Delta^{ij,kl} \dl_3(\bx)$ with
\begin{eqnarray}
       \Delta_{ij} &=& \dl_{ij} - \fr{\pd_i \pd_j}{\lap3} ,
              \nonumber \\
       \Delta_{ij,kl} &=& \half \left( \Delta_{ik} \Delta_{jl} + \Delta_{il} \Delta_{jk} 
                                       - \Delta_{ij} \Delta_{kl}  \right) .
\end{eqnarray}
These differential operators satisfy the conditions $\pd^i \Delta_{ij}=0$, $\Delta^j_{~j}=2$, $\pd^i \Delta_{ij,kl}=0$ and $\Delta^i_{~i,kl}=0$, and also satisfy the relations $\Delta_{ik} \Delta^k_{~j} = \Delta_{ij}$ and $\Delta_{ij,kl} \Delta^{kl,}_{~~~mn} = \Delta_{ij,mn}$.

The Hamiltonian is expressed as
\begin{eqnarray}
    H &=& \int d^3 \bx \l: \biggl\{  - \half \P^{ij}_\u \P^\u_{ij} + \P_\h^{ij} \u_{ij} + \u^{ij} \lap3 \u_{ij} 
              + \half \lap3 \h^{ij} \lap3 \h_{ij} 
                 \nonumber \\
      && \qquad\qquad\quad
           + \fr{1}{4} \P^j \invlap3 \P_j - \lap3 \h^j \lap3 \h_j
                           \biggr\} \r: ,
\end{eqnarray}
where $\invlap3 = 1/\lap3$.

The equations of motion of the transverse traceless tensor and the transverse vector modes are given by $\pd^4 \h^{ij}=0$ and $\lap3 \pd^2 \h^j=0$, which can be written in terms of the momentum variables as $\pd_\eta \P_\h^{ij}= -\dlap3 \h^{ij}$ and $\pd_\eta \P^j = 2\dlap3 \h^j$, respectively.

As in the case of the Riegert field, the transverse traceless tensor mode is expanded by $e^{ik_\mu x^\mu}$ and $\eta e^{ik_\mu x^\mu}$. Decomposing into the annihilation and creation parts as $\h^{ij} = \h^{ij}_< + \h^{ij}_>$, where $\h^{ij}_> = \h^{ij \dag}_<$, the annihilation part is expanded as
\begin{eqnarray}
    \h^{ij}_<(x) =  \int \fr{d^3 \bk}{(2\pi)^{3/2}} \fr{1}{2 \om^{3/2}}
                     \left\{ \sfc^{ij}(\bk) + i\om \eta \sfd^{ij}(\bk) \right\} e^{ik_\mu x^\mu} .
\end{eqnarray}
On the other hand, since the equation of motion of the transverse vector mode is the second order with respect to the time derivative, it is expanded as $\h^j=\h^j_< + \h^j_>$ with $\h^j_> = \h^{j \dag}_<$ and
\begin{eqnarray}
    \h^j_< (x) = \int \fr{d^3 \bk}{(2\pi)^{3/2}} \fr{1}{2 \om^{3/2}} \sfe_j(\bk)  e^{ik_\mu x^\mu} .
\end{eqnarray}
The annihilation parts of other variables are given by
\begin{eqnarray}
    \u^{ij}_< (x) &=& -i \int \fr{d^3 \bk}{(2\pi)^{3/2}} \fr{1}{2 \om^{1/2}}
                    \left\{ \sfc^{ij}(\bk) +( -1 +i\om \eta) \sfd^{ij}(\bk) \right\} e^{ik_\mu x^\mu} ,
                    \nonumber \\
    \P^{ij}_{\u <} (x) &=& \int \fr{d^3 \bk}{(2\pi)^{3/2}} \fr{\om^{1/2}}{2}
                   \left\{ \sfc^{ij}(\bk) +( -2 +i\om \eta) \sfd^{ij}(\bk) \right\} e^{ik_\mu x^\mu} ,
                    \nonumber \\
    \P^{ij}_{\h <}(x) &=& -i \int \fr{d^3 \bk}{(2\pi)^{3/2}} \fr{\om^{3/2}}{2}
                  \left\{ \sfc^{ij}(\bk) +( 1 +i\om \eta) \sfd^{ij}(\bk) \right\} e^{ik_\mu x^\mu} ,
                    \nonumber \\
    \P^j_<(x) &=& i \int \fr{d^3 \bk}{(2\pi)^{3/2}} \om^{3/2}
                   \sfe^j(\bk) e^{ik_\mu x^\mu} .
\end{eqnarray}

Substituting these expressions into the canonical commutation relations (\ref{equal-time commutation relation for tensor field}), we obtain the following commutation relations:
\begin{eqnarray}
      \left[ \sfc^{ij}(\bk), \sfc^{kl \dag}(\bk^\pp) \right] &=& \dl_3^{ij,kl}(\bk - \bk^\pp) ,
           \nonumber \\
      \left[ \sfc^{ij}(\bk), \sfd^{kl \dag}(\bk^\pp) \right] &=& \left[ \sfd^{ij}(\bk), \sfc^{kl \dag}(\bk^\pp) \right]
         = \dl_3^{ij,kl} (\bk -\bk^\pp),
             \nonumber \\
      \left[ \sfd^{ij} (\bk), \sfd^{kl \dag}(\bk^\pp) \right] &=& 0 ,
             \nonumber \\
      \left[ \sfe^i(\bk), \sfe^{j \dag}(\bk^\pp) \right] &=& -\dl_3^{ij}(\bk - \bk^\pp) , 
\end{eqnarray}
where the momentum representations of the delta functions $\dl_3^{ij}(\bk)$ and $\dl_3^{ij,kl}(\bk)$ are defined by multiplying the ordinary delta function $\dl_3(\bk)$ by the functions,
\begin{eqnarray}
    {\tilde \Delta}_{ij}(\bk) &=& \dl_{ij} - \fr{k_i k_j}{\bk^2} ,
          \nonumber \\
    {\tilde \Delta}_{ij,kl}(\bk) &=& \half \left\{ {\tilde \Delta}_{ik}(\bk) {\tilde \Delta}_{jl}(\bk) 
                         + {\tilde \Delta}_{il}(\bk) {\tilde \Delta}_{jk}(\bk) 
                         - {\tilde \Delta}_{ij}(\bk) {\tilde \Delta}_{kl}(\bk)  \right\} ,
\end{eqnarray}
respectively.

The commutation relations can be simplified by introducing the polarization vector $\veps_{(a)}^i ~(a=1,2)$ and the polarization tensor $\veps_{(a)}^{ij}~(a=1,2)$ that satisfy the transverse condition $k_i \veps^i_{(a)}=0$ and the transverse traceless conditions $k_i \veps_{(a)}^{ij}(\bk)= \veps^{~i}_{(a) i}(\bk)=0$, respectively. They are normalized as
\begin{eqnarray}
     \sum_{a=1}^2 \veps^i_{(a)}(\bk) \veps^j_{(a)}(\bk) = {\tilde \Delta}^{ij}(\bk) , \qquad
     \veps_{(a)}^j(\bk) \veps_{(b) j} (\bk) = \dl_{ab} 
\end{eqnarray}
and
\begin{eqnarray}
     \sum_{a=1}^2 \veps^{ij}_{(a)}(\bk) \veps^{kl}_{(a)}(\bk) = {\tilde \Delta}^{ij,kl}(\bk) , \qquad
    \veps_{(a)}^{ij}(\bk) \veps_{(b) ij}(\bk) = \dl_{ab} .
\end{eqnarray}
Each mode is then expanded as
\begin{eqnarray}
     \sfc^{ij}(\bk) &=& \sum_{a=1}^2 \veps_{(a)}^{ij}(\bk) \sfc_{(a)}(\bk),  \quad 
     \sfd^{ij}(\bk) = \sum_{a=1}^2 \veps_{(a)}^{ij}(\bk) \sfd_{(a)}(\bk), 
           \nonumber \\
     \sfe^j(\bk) &=& \sum_{a=1}^2 \veps_{(a)}^j(\bk) \sfe_{(a)}(\bk) .
\end{eqnarray}
The commutation relations are simply given by
\begin{eqnarray}
      \left[ \sfc_{(a)}(\bk), \sfc^\dag_{(b)}(\bk^\pp) \right] &=&  \dl_{ab} \dl_3 (\bk - \bk^\pp) ,
           \nonumber \\
      \left[ \sfc_{(a)}(\bk), \sfd^\dag_{(b)}(\bk^\pp) \right] &=& \left[ \sfd_{(a)}(\bk), \sfc^\dag_{(b)}(\bk^\pp) \right]
         = \dl_{ab} \dl_3 (\bk -\bk^\pp),
             \nonumber \\
      \left[ \sfd_{(a)} (\bk), \sfd^\dag_{(b)}(\bk^\pp) \right] &=& 0 ,  
             \nonumber \\
      \left[ \sfe_{(a)} (\bk), \sfe^\dag_{(b)}(\bk^\pp) \right] &=& - \dl_{ab} \dl_3(\bk-\bk^\pp)
\end{eqnarray}
The Hamiltonian is expressed as
\begin{equation}
     H = \sum_{a=1}^2 \int d^3 \bk \om \left\{ \sfc_{(a)}^\dag(\bk) \sfd_{(a)}(\bk) + \sfd^\dag_{(a)}(\bk) \sfc_{(a)}(\bk) 
                                  - 2 \sfd^\dag_{(a)}(\bk) \sfd_{(a)}(\bk) 
                                  - \sfe^\dag_{(a)}(\bk) \sfe_{(a)}(\bk) \right\} .
\end{equation}

Finally, we calculate the two-point functions of the transverse traceless tensor and transverse vector modes. We here introduce the hermitian fields $H^{(a)}$ and $Y^{(a)}$ with the annihilation parts defined by
\begin{eqnarray}
    H^{(a)}_<(x) &=&  \int \fr{d^3 \bk}{(2\pi)^{3/2}} \fr{1}{2 \om^{3/2}}
                  \left\{ \sfc_{(a)}(\bk) + i\om \eta \sfd_{(a)}(\bk) \right\} e^{ik_\mu x^\mu} ,
              \nonumber \\
    Y^{(a)}_< (x) &=& \int \fr{d^3 \bk}{(2\pi)^{3/2}} \fr{1}{2 \om^{3/2}}
                         \sfe_{(a)}(\bk)  e^{ik_\mu x^\mu} .
\end{eqnarray}
We then obtain the two-point function $\lang H^{(a)}(x) H^{(b)}(x^\pp) \rang = \dl_{ab} \lang H(x) H(x^\pp) \rang$, where
\begin{eqnarray}
      \lang H(x) H(x^\pp) \rang 
        &=& -\fr{1}{16\pi^2} \log \left\{ \left[ -(\eta - \eta^\pp -i\eps)^2 + (\bx - \bx^\pp)^2 \right]  
                                               z^2 e^{2\gm-2} \right\} 
              \nonumber \\
       && -\fr{1}{16\pi^2}\fr{i\eps}{|\bx - \bx^\pp|}
            \log \fr{\eta - \eta^\pp -i\eps - |\bx - \bx^\pp|}{\eta - \eta^\pp -i\eps+|\bx - \bx^\pp|} .
\end{eqnarray}
And also, we obtain $\lang Y^{(a)}(x) Y^{(b)}(x^\pp) \rang = \dl_{ab} \lang Y(x) Y(x^\pp) \rang$, where 
\begin{eqnarray}
      \lang Y(x) Y(x^\pp) \rang 
        &=& \fr{1}{16\pi^2} \log \left\{ \left[ -(\eta - \eta^\pp -i\eps)^2 + (\bx - \bx^\pp)^2 \right]  
                                               z^2 e^{2\gm-2} \right\} 
              \nonumber \\
       && - \fr{1}{16\pi^2}\fr{\eta-\eta^\pp-i\eps}{|\bx - \bx^\pp|}
            \log \fr{\eta - \eta^\pp -i\eps - |\bx - \bx^\pp|}{\eta - \eta^\pp -i\eps+|\bx - \bx^\pp|} .
\end{eqnarray}
Using these functions, we obtain the following expressions:
\begin{eqnarray}
   \lang \h_{ij}(x) \h_{kl}(x^\pp) \rang &=& \Delta_{ij,kl}(\bx) \lang H(x) H(x^\pp) \rang ,
           \nonumber \\
   \lang \h^i(x) \h^j(x^\pp) \rang &=& \Delta^{ij}(\bx) \lang Y(x) Y(x^\pp) \rang . 
\end{eqnarray}

\subsection{Generators of Conformal Symmetry}
\noindent

As stressed in section 2, the transverse traceless tensor mode $\h_{ij}$ and the transverse vector mode $\h_j$ themselves are not gauge invariant because they mix under the residual gauge transformation, or the conformal transformation (\ref{conformal transformation}).

We here write out the generator of the transformation in the radiation gauge. The stress tensor of the traceless tensor field is derived from the Weyl action in Appendix E.  According to the definition (\ref{definition of generator}), we obtain the following expressions. The generators of translations are given by
\begin{eqnarray}
     P_0 &=& H = \int \d3x \calA ,
              \nonumber \\
     P_j &=& \int \d3x \calB_j ,
\end{eqnarray}
where
\begin{eqnarray}
    \calA &=& - \half \l: \P^{kl}_\u \P^\u_{kl} \r: + \l: \P_\h^{kl} \u_{kl} \r: + \l: \u^{kl} \lap3 \u_{kl} \r: 
              + \half \l: \lap3 \h^{kl} \lap3 \h_{kl} \r: 
                 \nonumber \\
           && + \fr{1}{4} \l: \P^k \invlap3 \P_k \r: - \l: \lap3 \h^k \lap3 \h_k \r: ,
                 \nonumber \\
    \calB_j &=& \l: \P_\u^{kl} \pd_j \u_{kl} \r:  + \l: \P_\h^{kl} \pd_j \h_{kl} \r:  + \l: \P^k \pd_j \h_k \r: .
\end{eqnarray}
The generators of Lorentz transformations are 
\begin{eqnarray}
    M_{0j} &=& \int \d3x \left\{ -\eta \calB_j - x_j \calA - \calC_j \right\} ,
                  \nonumber \\
    M_{ij} &=& \int \d3x \left\{  x_i \calB_j - x_j \calB_i + \calC_{ij} \right\} ,
\end{eqnarray}
where
\begin{eqnarray}
    \calC_j &=& \l: \P_\u^{kl} \pd_j \h_{kl} \r: + \l: \P^k_{\u ~j} \invlap3 \P_k \r:  + 2 \l: \P^k_{\h ~j} \h_k \r:
                + \l: \h^k_{~j} \P_k \r:   + 2 \l: \u^k_{~j} \lap3 \h_k \r: ,
                 \nonumber \\
    \calC_{ij} &=& 2 \left( \l: \P^k_{\u ~i} \u_{kj} \r: - \l: \P^k_{\u ~j} \u_{ki} \r: \right)
                   + 2 \left( \l: \P^k_{\h ~i} \h_{kj} \r: - \l: \P^k_{\h ~j} \h_{ki} \r: \right) 
                                    \nonumber \\
               &&  + \l: \P_i \h_j \r: - \l: \P_j \h_i \r: .
\end{eqnarray}
The generator of dilatation is
\begin{eqnarray}
    D = \int \d3x \left\{ \eta \calA + x^k \calB_k + \l: \P_\u^{kl} \u_{kl} \r: \right\} .
\end{eqnarray}
The generators of special conformal transformations are
\begin{eqnarray}
   K_0 &=& -\eta^2 P_0 + 2\eta D + N_0 ,
          \nonumber \\
   K_j &=& \eta^2 P_j + 2 \eta M_{0j} + N_j ,
\end{eqnarray}
where the operators defined by $N_0=\int \d3x \bx^2 T_{00}$ and $N_j=\int \d3x ( \bx^2 T_{0j} -2x_j x^k T_{0k} )$ are given by
\begin{eqnarray}
    N_0 &=& \int \d3x \biggl\{  \bx^2 \calA    + 2 x^k \calC_k  
              - 2 \l: \u^{kl} \u_{kl} \r:  - \l: \pd^m \h^{kl} \pd_m \h_{kl} \r:  
                 \nonumber \\          
         && \qquad\qquad
            - \fr{5}{4} \l: \invlap3 \P^k \invlap3 \P_k \r: - 4 \l: \pd^k \h^l \pd_k \h_l \r: 
             \biggr\} ,
                \nonumber \\
    N_j &=& \int \d3x \Bigl\{ \bx^2 \calB_j - 2 x_j x^k \calB_k  + 2x^k \calC_{kj}
            - 2 x_j \l: \P_\u^{kl} \u_{kl} \r:   
                \nonumber \\
         && \qquad\qquad
            - 2 \l: \u^{kl} \pd_j \h_{kl} \r:    + 2 \l: \invlap3 \P^k \pd_j \h_k \r:    
            - 4 \l: \P^k_{\u ~j} \h_k \r:        
                \nonumber \\
         && \qquad\qquad
            - 4 \l: \u^k_{~j} \invlap3 \P_k \r:     + 4 \l: \h^k_{~j} \lap3 \h_k \r: 
            \Bigr\} .
\end{eqnarray}

In this paper, we do not check that these generators form the closed algebra of conformal symmetry on $M^4$. On the other hand, on the $R \times S^3$ background, we have already shown that the corresponding generators form the closed algebra quantum mechanically \cite{hh}.

\subsection{Transformation Laws and Fradkin-Palchik Terms}
\noindent

In this subsection, we study the transformation property of the field. First, we write down the expressions of the transformations $\dl_\zeta h_{\mu\nu}$ in (\ref{conformal transformation}) in the radiation gauge, which are given by
\begin{eqnarray}
    \dl_\zeta \h_{rs} &=& \zeta^\lam \pd_\lam \h_{rs}  + \h_{rk} \pd_s \zeta^k  + \h_{sk} \pd_r \zeta^k 
           + \h_r \pd_s \zeta^0  + \h_s \pd_r \zeta^0  - \half \h_{rs} \pd_\lam \zeta^\lam ,
          \nonumber \\
    \dl_\zeta \h_r &=& \zeta^\lam \pd_\lam \h_r + \h_r \pd_\eta \zeta^0 + \h_s \pd_r \zeta^s  
                       + \h_{rs} \pd_\eta \zeta^s  - \half \h_r \pd_\lam \zeta^\lam .
             \label{transformation in radiation gauge}
\end{eqnarray}
These transformations, however, do not preserve the gauge-fixing conditions (\ref{radiation gauge}). This fact is known to be a general feature of conformal transformations that appears in gauge theories.

As discussed by Fradkin and Palchik \cite{fp, kluwer}, the violation of invariance can be compensated by a certain gauge transformation whose parameters depend on the field. In this case, it is expressed as
\begin{eqnarray}
   i \left[ Q_\zeta, h_{\mu\nu} \right] = \dl_\zeta h_{\mu\nu} + \dl_\tkappa h_{\mu\nu} ,
           \label{tranformation law with Fradkin-Palchik term}
\end{eqnarray}
where the second term in the right-hand side, called the Fradkin-Palchik term, has the form of the gauge transformation $\dl_\tkappa h_{\mu\nu} = \pd_\mu \tkappa_\nu + \pd_\nu \tkappa_\mu - \eta_{\mu\nu} \pd_\lam \tkappa^\lam/2$ with the nonlocal field-dependent gauge parameter $\tkappa^\mu$. Here, using the generators given above, we demonstrate this transformation law and determine $\tkappa^\mu$ in the radiation gauge.

For $P_\mu$ of translations, $D$ of dilatations and $M_{ij}$ of spatial rotations, the transformation laws of the transverse traceless tensor mode are calculated using the canonical commutation relations (\ref{equal-time commutation relation for tensor field}) as
\begin{eqnarray}
  i \left[ P_\mu, \h_{rs} \right] &=& \pd_\mu \h_{rs}, 
             \nonumber \\ 
  i \left[ D, \h_{rs} \right] &=& x^\lam \pd_\lam \h_{rs} ,
             \nonumber \\
  i \left[ M_{ij} , \h_{rs} \right] &=& \left( x_i \pd_j - x_j \pd_i \right) \h_{rs} 
                + \dl_{ri} \h_{sj} - \dl_{rj} \h_{si} 
                + \dl_{si} \h_{rj} - \dl_{sj} \h_{ri}
\end{eqnarray}
and those of the transverse vector mode are
\begin{eqnarray}
  i \left[ P_\mu, \h_r \right] &=& \pd_\mu \h_r,  
             \nonumber \\
  i \left[ D, \h_r \right] &=& x^\lam \pd_\lam \h_r ,
             \nonumber \\
  i \left[ M_{ij} , \h_r \right] &=& \left( x_i \pd_j - x_j \pd_i \right) \h_r 
                + \dl_{ri} \h_j  - \dl_{rj} \h_i .
\end{eqnarray}
These equations are nothing but the transformations (\ref{transformation in radiation gauge}) and the Fradkin-Palchik term does not appear here.

For $M_{0j}$ of Lorentz boosts and $K_\mu$ of special conformal transformations, we find that the Fradkin-Palchik term comes out in order to preserve the gauge-fixing condition, which are given as follows:
\begin{eqnarray}
  i \left[ M_{0j} , \h_{rs} \right] &=& \left( -\eta \pd_j - x_j \pd_\eta \right) \h_{rs} 
                - \dl_{rj} \h_s  - \dl_{sj} \h_r  
             \nonumber \\
               && + \pd_r \left( \tkappa^B_s \right)_j + \pd_s \left( \tkappa^B_r \right)_j 
                  - \half \dl_{rs} \pd_\lam \left( \tkappa^\lam_B \right)_j,
             \nonumber \\   
  i \left[ K_0 , \h_{rs} \right] &=& \left( \eta^2 + \bx^2 \right) \pd_\eta \h_{rs} 
                 + 2 \eta x^k \pd_k \h_{rs}   + 2 x_r \h_s  + 2 x_s \h_r  
             \nonumber \\ 
               && + \pd_r \left( \tkappa^S_s \right)_0 + \pd_s \left( \tkappa^S_r \right)_0 
                  - \half \dl_{rs} \pd_\lam \left( \tkappa^\lam_S \right)_0,
             \nonumber \\
  i \left[ K_j , \h_{rs} \right] &=& \left( - \eta^2 + \bx^2 \right) \pd_j \h_{rs} 
                 - 2 x_j x^k \pd_k \h_{rs} - 2 \eta x_j \pd_\eta \h_{rs}   
                 + 2 x_r \h_{sj}  + 2 x_s \h_{rj}   
             \nonumber \\
                && - 2\dl_{rj} x^k \h_{ks}  - 2\dl_{sj} x^k \h_{kr}
                    - 2\eta \dl_{rj} \h_s  - 2 \eta \dl_{sj} \h_r 
             \nonumber \\ 
               && + \pd_r \left( \tkappa^S_s \right)_j + \pd_s \left( \tkappa^S_r \right)_j 
                  - \half \dl_{rs} \pd_\lam \left( \tkappa^\lam_S \right)_j                
\end{eqnarray}
and 
\begin{eqnarray}
  i \left[ M_{0j} , \h_r \right] &=& \left( -\eta \pd_j - x_j \pd_\eta \right) \h_r  - \h_{rj}  
                 + \pd_\eta \left( \tkappa^B_r \right)_j  + \pd_r \left( \tkappa^B_0 \right)_j  ,
             \nonumber \\   
  i \left[ K_0 , \h_r  \right] &=& \left( \eta^2 + \bx^2 \right) \pd_\eta \h_r 
                 + 2 \eta x^k \pd_k \h_r   + 2 x^k \h_{kr}  
                 + \pd_\eta \left( \tkappa^S_r \right)_0  + \pd_r \left( \tkappa^S_0 \right)_0 ,
             \nonumber \\
  i \left[ K_j , \h_r \right] &=& \left( - \eta^2 + \bx^2 \right) \pd_j \h_r 
                 - 2 x_j x^k \pd_k \h_r - 2 \eta x_j \pd_\eta \h_r  + 2 x_r \h_j   
             \nonumber \\
                &&  - 2\dl_{rj} x^k \h_k    - 2\eta \h_{rj}
                    + \pd_\eta \left( \tkappa^S_r \right)_j  + \pd_r \left( \tkappa^S_0 \right)_j ,               
\end{eqnarray}
where the nonlocal field-dependent gauge parameters $(\tkappa_B^\lam)_j$ for Lorentz boosts are given by
\begin{eqnarray}
    \left( \tkappa^0_B \right)_j &=& -\fr{3}{2} \invlap3 \pd_\eta \h_j ,
           \nonumber \\
    \left( \tkappa^r_B \right)_j &=& \invlap3 \pd_\eta \h^r_{~j} + \invlap3 \pd_j \h^r 
                   -\half \invlap3 \pd^r \h_j       
\end{eqnarray}
and $(\tkappa_S^\lam)_\mu$ for special conformal transformations are given by
\begin{eqnarray}
    \left( \tkappa^0_S \right)_0 &=& 3 \invlap3 \pd_\eta \left( x^k \h_k \right) ,
           \nonumber \\
    \left( \tkappa^r_S \right)_0 &=& -2 \invlap3 \pd_\eta \left( x^k \h^r_{~k} \right)
                  - 2 \invlap3 \pd_k \left( x^k \h^r \right)  
                  - 2 \invlap3 \h^r     + \invlap3 \pd^r \left( x^k \h_k \right), 
           \nonumber \\
    \left( \tkappa^0_S \right)_j &=& - 3 \invlap3 \pd_\eta \left( \eta \h_j \right) + 6 \invlap3 \h_j,
           \nonumber \\
    \left( \tkappa^r_S \right)_j &=& 2 \invlap3 \pd_\eta \left( \eta \h^r_{~j} \right) 
                   - 8 \invlap3 \h^r_{~j}     + 2 \invlap3 \pd_j \left( \eta \h^r \right) 
                   - \invlap3 \pd^r \left( \eta \h_j \right) .
\end{eqnarray}

Here, we show that the Fradkin-Palchik term is indeed necessary to preserve the gauge-fixing condition. Let us consider the $(00)$ component of (\ref{tranformation law with Fradkin-Palchik term}). The left-hand side trivially vanishes due to the gauge-fixing condition $h_{00}=0$, while the right-hand side is nontrivial because $\dl_\zeta h_{00} = 2 \h_r \pd_\eta \zeta^r$, which survives in the cases of Lorentz boosts and special conformal transformations. The Fradkin-Palchik term $\dl_\tkappa h_{00}$ just cancels it. The transverse conditions are also preserved.

\section{BRST Operator}
\setcounter{equation}{0}
\noindent

Finally, we rewrite the quantum diffeomorphism (\ref{conformal transformation}) and the physical field condition (\ref{physical field condition}) in the BRST formalism \cite{kato, brs, ko}.

The BRST transformation is defined by replacing $\zeta^\mu$ in the gauge transformations (\ref{conformal transformation}) with the corresponding gauge ghost $c^\mu$. It is expanded by 15 Grassmann modes as
\begin{eqnarray}
   c^\lam(x) &=& \rc_-^\mu \left( \zeta_T^\lam \right)_\mu  + \rc^{\mu\nu} \left( \zeta_L^\lam \right)_{\mu\nu}
               + \rc \zeta_D^\lam  + \rc_+^\mu \left( \zeta_S^\lam \right)_\mu
                  \nonumber \\
             &=& \rc_-^\lam   + 2 x_\mu \rc^{\mu\lam}  + x^\lam \rc   + x^2 \rc_+^\lam  - 2 x^\lam x_\mu \rc_+^\mu ,
\end{eqnarray}
where $\rc_-^\lam$, $\rc^{\mu\nu}$, $\rc$, $\rc_+^\lam$ are hermitian operators and $\rc^{\mu\nu}$ is antisymmetric. The $\rc$ and $\rc^{\mu\nu}$ modes have no dimensions, while $\rc_-^\mu$ and $\rc_+^\mu$ have dimensions $-1$ and $1$, respectively.

We also introduce the 15 antighosts $\rb_-^\lam$, $\rb^{\mu\nu}$, $\rb$ and $\rb_+^\lam$ with the same properties that the gauge ghosts have. The anticommutation relations between gauge ghosts and antighosts are set as
\begin{eqnarray}
   \left\{ \rc , \rb \right\} &=& 1 , 
        \nonumber \\
    \left\{ \rc^{\mu\nu} , \rb^{\lam\s} \right\} &=& \eta^{\mu\lam} \eta^{\nu\s} - \eta^{\mu\s} \eta^{\nu\lam} , 
        \nonumber \\
   \left\{ \rc_-^\mu , \rb_+^\nu \right\} &=& \left\{ \rc_+^\mu , \rb_-^\nu \right\} = \eta^{\mu\nu} .
\end{eqnarray}
From these algebra, we find that the generators of conformal algebra in the gauge ghost sector represented by the abbreviation ``$\gh$" are given by\footnote{ 
These expressions can be determined by imposing the condition that the ghost field $c^\mu$ transforms as a vector field with conformal dimension $-1$. 
}  
\begin{eqnarray}
    P_\gh^\mu &=& i \left( - 2 \rb \rc_+^\mu  + \rb_+^\mu \rc   + \rb^\mu_{~\lam} \rc_+^\lam  
                           + 2 \rb_+^\lam \rc^\mu_{~\lam}   \right) ,
                    \nonumber \\
    M_\gh^{\mu\nu} &=& i \left( \rb_+^\mu \rc_-^\nu  - \rb_+^\nu \rc_-^\mu   
                                + \rb_-^\mu \rc_+^\nu  - \rb_-^\nu \rc_+^\mu
                                + \rb^{\mu\lam} \rc^\nu_{~\lam}  - \rb^{\nu\lam} \rc^\mu_{~\lam}  \right) ,
                    \nonumber \\
    D_\gh &=& i \left( \rb_-^\lam \rc_{+ \lam}   - \rb_+^\lam \rc_{- \lam}  \right), 
                    \nonumber \\
    K^\mu_\gh &=& i \left(  2 \rb \rc_-^\mu  - \rb_-^\mu \rc   + \rb^\mu_{~\lam} \rc_-^\lam  
                           + 2 \rb_-^\lam \rc^\mu_{~\lam}   \right) . 
\end{eqnarray}

Using these generators, the BRST operator is defined by 
\begin{eqnarray}
     Q_\BRST &=& \rc_-^\mu \left( P_\mu + \half P^\gh_\mu \right) 
               + \rc^{\mu\nu} \left( M_{\mu\nu} + \half M^\gh_{\mu\nu} \right) 
               + \rc \left( D + \half D^\gh \right) 
                  \nonumber \\
             &&  + \rc_+^\mu \left( K_\mu + \half K^\gh_\mu \right) 
                  \nonumber \\
             &=& \rc \left( D + D^\gh \right) + \rc^{\mu\nu} \left( M_{\mu\nu} + M^\gh_{\mu\nu} \right)
              - \rb N - \rb^{\mu\nu} N_{\mu\nu} + {\hat Q} ,
\end{eqnarray}
where $P_\mu$, $M_{\mu\nu}$, $D$ and $K_\mu$ represent the sum of the generators of conformal algebra other than the gauge ghost sector. The other operators are defined by
\begin{eqnarray}
    N &=& 2 i \rc_+^\mu \rc_{- \mu} ,
            \nonumber \\
    N^{\mu\nu} &=& \fr{i}{2} \left( \rc_+^\mu \rc_-^\nu + \rc_-^\mu \rc_+^\nu \right) 
                   + i \rc^{\mu\lam} \rc^\nu_{~\lam} ,
            \nonumber \\
    {\hat Q} &=& \rc_-^\mu P_\mu + \rc_+^\mu K_\mu .
\end{eqnarray}
The nilpotency can be shown by using the conformal algebra for $P_\mu$, $M_{\mu\nu}$, $D$ and $K_\mu$ as
\begin{eqnarray}
   Q_\BRST^2 = {\hat Q}^2 - N D - 2i \rc_+^\mu \rc_-^\nu M_{\mu\nu} = 0.
\end{eqnarray}

Here, the anticommutation relations between the BRST operator and the antighosts are computed as
\begin{eqnarray}
   \left\{ Q_\BRST, \rb \right\} &=& D + D_\gh , 
           \nonumber \\
   \left\{ Q_\BRST, \rb^{\mu\nu} \right\} &=& 2 \left( M^{\mu\nu} + M_\gh^{\mu\nu} \right) ,
           \nonumber \\
   \left\{ Q_\BRST, \rb_-^\mu \right\} &=& K^\mu + K_\gh^\mu , 
           \nonumber \\
   \left\{ Q_\BRST, \rb_+^\mu \right\} &=& P^\mu + P_\gh^\mu .
\end{eqnarray}
Therefore, the nilpotency can be also expressed as $[Q_\BRST, D+D_\gh ]=0$ and so on.

In terms of the BRST operator, the physical field conditions (\ref{physical field condition}) are now written by the single equation,
\begin{eqnarray}
    \left[ Q_\BRST, \int d^4x {\cal O}(x) \right] = 0.
\end{eqnarray}

Now, the results found in section 4 are summarized as follows. The conformal transformations for Riegert and gauge ghost fields are expressed as
\begin{eqnarray}
    i \left[ Q_\BRST , \phi(x) \right] &=& c^\mu\pd_\mu \phi(x) + \fr{1}{4} \pd_\mu c^\mu(x) ,
               \nonumber \\
    i \left\{ Q_\BRST , c^\mu(x) \right\} &=& c^\nu \pd_\nu c^\mu(x) .               
\end{eqnarray}
The BRST transformation of the exponential operator $V_\a$ (\ref{conformal field V_a}) is given by 
\begin{eqnarray}
    i \left[ Q_\BRST , V_\a(x) \right] = c^\mu\pd_\mu V_\a (x) + \fr{h_\a}{4} \pd_\mu c^\mu V_\a(x) .
\end{eqnarray}
Thus, we obtain
\begin{eqnarray}
    i \left[ Q_\BRST, \int d^4 x V_\a(x) \right] = \int d^4 x \pd_\mu \{ c^\mu V_\a(x) \} = 0 
\end{eqnarray}
for $h_\a=4$.

\section{Conclusion}
\setcounter{equation}{0}
\noindent

We have studied four dimensional quantum gravity in a nonperturbative manner capable of describing spacetime dynamics beyond the Planck scale. The model has been constructed on the basis of conformal gravity incorporating the Riegert-Wess-Zumino action that is required to preserve the diffeomorphism invariance. Indeed, we have found that quantum diffeomorphism becomes complete in the combined system of Riegert-Wess-Zumino and Weyl actions.

The model has been described as a certain conformal field theory defined on a background spacetime that can be chosen arbitrary so long as it is conformally flat due to the background free nature. In this paper, we studied the model by employing the Minkowski background $M^4$ in practice.

The generator of conformal transformation was constructed on $M^4$ and the transformation law of the field was studied to clarify the physical property. Since conformal invariance originates from diffeomorphism invariance, physical fields must be invariant under the conformal transformation. Thus, conformal fields themselves are not physical fields. We found that physical fields are given by diffeomorphism invariant fields that are described by spacetime volume integrals of scalar fields with conformal dimension 4, and thus tensor fields outside the unitarity bound 
are excluded.\footnote{ 
The Weyl tensor, for example, has spin $s=2$ and conformal domension $d=2$, which lies outside the unitarity bound $d \geq 2+s$ \cite{kluwer}. It is excluded by the physical condition, but its squared becomes physical. 
} 
Thus, conformal invariance from diffeomorphism invariance gives more strong conditions to the field than those in usual conformal field theories.

We also constructed the nilpotent BRST operator for quantum diffeomorphism. The physical field condition was represented in terms of the BRST operator.

The results were confirmed to be consistent with those studied on the curved background $R \times S^3$. The advantages of using the $M^4$ background are that it is, of course, most familiar and also field operators and their correlation functions become simple compared with those on the curved background, which are helpful in future developments.

Finally, we give a comment on how to resolve the ghost problem at a higher order of the coupling $t$. In this case, we can apply an old idea by Tomboulis \cite{tomboulis} based on Lee and Wick's works \cite{lw} in the 1970s. The essence of their idea is that in the full propagator, including radiative corrections, the ghost pole moves to a complex plane so that the ghost mode does not appear in the real world. However, this idea has a problem that the ghost mode appears as a real pole when interactions turn off at the UV limit. In our model, the conformal symmetry just compensates for this weak point.


\vspace{1cm}

\appendix

\section{Wess-Zumino Condition and Background Metric Independence}
\setcounter{equation}{0}
\noindent

The Riegert-Wess-Zumino action (\ref{Riegert action}) can be written in the integral form,\footnote{ 
As a comparison, the Liouville-Polyakov action in two dimensions is given by $S_{\rm L}= -(b_{\rm L}/4\pi)\int d^2 x \int^\phi_0 d\phi \sq{-g}R = -(b_{\rm L}/4\pi)\int d^2 x \sq{-\hg}( \phi \hDelta_2 \phi + \hR \phi)$, where $\Delta_2= -\nb^2$ and $b_{\rm L}=(25-c)/6$ for two dimensional quantum gravity coupled to conformal field theory with central charge $c$ \cite{polyakov, kpz, dk}.
} 
\begin{eqnarray}
     S_{\rm RWZ}(\phi,\hg) = - \fr{b_1}{(4\pi)^2} \int d^4 x \int^\phi_0 d \phi \sq{-g} E_4 ,
\end{eqnarray}
where $E_4 = G_4 - 2 \nb^2 R /3$ is the modified Euler density and $G_4= R_{\mu\nu\lam\s}^2 -4R_{\mu\nu}^2 +R^2$ is the usual Euler density. The integral can be easily carried out using the relation $\sq{-g}E_4 = \sq{-\hg}(4\hDelta_4 \phi + \hE_4)$. Here, the metric field $g_{\mu\nu}$ is taken to be $e^{2\phi}\hg_{\mu\nu}$ given at the vanishing coupling limit, where $\hg_{\mu\nu}$ is conformally flat. The Weyl action is then described by the kinetic term of $h_{\mu\nu}$ defined on the metric field $g_{\mu\nu}$.

By dividing the range of integration $[0,\phi]$ into $[0,\om]$ and $[\om,\phi]$, we find that the action satisfies the Wess-Zumino integrability condition \cite{wz, bcr, riegert}
\begin{eqnarray}
    S_{\rm RWZ}(\phi, \hg) = S_{\rm RWZ}(\om,\hg) + S_{\rm RWZ}(\phi-\om,e^{2\om}\hg) .
        \label{Wess-Zumino condition}
\end{eqnarray}
This condition guarantees that the path integral of Riegert field can be carried out exactly in a manner independent of how to choose the path.

Background free nature just comes out as a consequence of performing the path integral over the Riegert field exactly. It is shown as follows:\footnote{ 
The UV divergences, which are proportional to ${\hat C}_{\mu\nu\lam\s}^2$ and $\hE_4$, vanish for a conformally flat background. 
} 
\begin{eqnarray}
   Z|_{e^{2\om}\hg} &=& \int [d\phi dh]_{e^{2\om}\hg} \exp \left\{ 
                iS_{\rm RWZ}(\phi,e^{2\om}\hg)+iI(e^{2\om}g) \right\}
          \nonumber \\
       &=& \int [d\phi dh]_\hg \exp \left\{ iS_{\rm RWZ}(\om,\hg) + iS_{\rm RWZ}(\phi,e^{2\om}\hg)
                                            + iI(e^{2\om}g) \right\}
          \nonumber \\
       &=& \int [d\phi dh]_{\hg} \exp \left\{ iS_{\rm RWZ}(\om,\hg)+
                iS_{\rm RWZ}(\phi-\om,e^{2\om}\hg)+iI(g) \right\}
           \nonumber \\
       &=& Z|_\hg .
\end{eqnarray}
In the second equality, we take into account the Jacobian factor $\exp\{iS_{\rm RWZ}(\om,\hg)\}$ arising when we rewrite the path integral measure. The third equality is derived by changing the variable of integration as $\phi \to \phi-\om$, considering that the measure $[d\phi]_\hg$ is invariant under such a local shift. In the last equality, we use the Wess-Zumino integrability condition (\ref{Wess-Zumino condition}).

Besides the Euler density, the square of Weyl tensor is also integrable. It produces another Wess-Zumino action $\sq{-g}\phi C_{\mu\nu\lam\s}^2$ accompanied with the beta function, but it disappears at the UV limit of $t=0$. The Ricci scalar $\sq{-g}R$ and the cosmological constant term $\sq{-g}$ are also integrable, while $\sq{-g}R^2$ is not integrable and thus it is discarded from the action $I$ to guarantee the integrability of the Riegert field. It is consistent with the fact that the generator of quantum diffeomorphism becomes complete in the system without $R^2$ action.

\section{Integral Formulae}
\setcounter{equation}{0}
\noindent

The following Fourier integral is useful: 
\begin{eqnarray}
    I_n(\eta,\bx) &=& \int_{\om >z} \fr{d^3 \bk}{(2\pi)^3} \fr{1}{\om^n} e^{-i\om(\eta -i\eps)+i\bk \cdot \bx}
              \nonumber \\
        &=& \fr{1}{(2\pi)^3} \int^\infty_z \om^2 d\om \int^1_{-1} d\cos\theta \int^{2\pi}_0 d\varphi
                \fr{1}{\om^n} e^{i\om|\bx|\cos\theta}e^{-i\om(\eta-i\eps)}
              \nonumber \\
        &=& \fr{1}{2\pi^2} \fr{1}{|\bx|} \int^\infty_z d\om \fr{1}{\om^{n-1}}\sin(\om|\bx|)e^{-i\om(\eta-i\eps)},
\end{eqnarray}
where $\om=|\bk|$. The parameters $\eps$ and $z$ are the UV and IR cutoffs, respectively. The integration is carried out under the condition $z \ll 1$, while $\eps$ is finite. The integral satisfies the relation $I_n(\eta,\bx) = i \pd_\eta I_{n+1}(\eta,\bx)$.

Here, we write down the results for several integers $n$ in the following:
\begin{eqnarray}
     I_3(\eta,\bx) &=& \fr{1}{4\pi^2} \biggl\{ - \log \left[ -(\eta-i\eps)^2 + \bx^2 \right]  
                                               - \log z^2e^{2\gm-2} 
                          +  \fr{\eta-i\eps}{|\bx|}\log \fr{\eta-i\eps-|\bx|}{\eta-i\eps+|\bx|}
                               \biggr\} ,
                 \nonumber \\
     I_2(\eta,\bx) &=& i \fr{1}{4\pi^2} \fr{1}{|\bx|} \log \fr{\eta-i\eps-|\bx|}{\eta-i\eps+|\bx|} ,
                 \nonumber \\
     I_1(\eta,\bx) &=& \fr{1}{2\pi^2} \fr{1}{-(\eta-i\eps)^2+\bx^2} ,
                 \nonumber \\
     I_0(\eta,\bx) &=& i \fr{1}{\pi^2} \fr{\eta-i\eps}{[-(\eta-i\eps)^2+\bx^2]^2} ,
                 \nonumber \\
     I_{-1}(\eta,\bx) &=& -\fr{1}{\pi^2} \fr{3(\eta-i\eps)^2+\bx^2}{[-(\eta-i\eps)^2+\bx^2]^3} ,
                 \nonumber \\
     I_{-2}(\eta,\bx) &=& -i\fr{12}{\pi^2} \fr{(\eta-i\eps)[(\eta-i\eps)^2+\bx^2]}{[-(\eta-i\eps)^2+\bx^2]^4} ,     
                 \nonumber \\
     I_{-3}(\eta,\bx) &=& \fr{12}{\pi^2} \fr{5(\eta-i\eps)^4+10(\eta-i\eps)^2\bx^2+\bx^4}{[-(\eta-i\eps)^2+\bx^2]^5} ,
                 \nonumber \\
     I_{-4}(\eta,\bx) &=& i\fr{120}{\pi^2} \fr{(\eta-i\eps)[3(\eta-i\eps)^4+10(\eta-i\eps)^2\bx^2+3\bx^4]}{[-(\eta-i\eps)^2+\bx^2]^6} .          
\end{eqnarray}

\section{Various OPE singularities}
\setcounter{equation}{0}
\noindent

We here write down the singular parts of the OPEs at the equal time used in sections 4 and 5. For four canonical variables, we obtain
\begin{eqnarray}
    \lang \phi(\bx) \phi(\bx^\pp) \rang &=& 
         -\fr{1}{4b_1} \log \left\{ \left[ (\bx -\bx^\pp)^2 + \eps^2 \right] z^2 e^{2\gm} \right\} 
         + \fr{1}{2b_1} \left[ 1 - h(\bx-\bx^\pp) \right] ,
              \nonumber \\
    \lang \phi(\bx) \chi(\bx^\pp) \rang &=& \lang \P_\chi(\bx) \P_\phi(\bx^\pp) \rang = 0,          
             \nonumber \\
    \lang \chi(\bx) \chi(\bx^\pp) \rang &=&  
          -\fr{1}{2b_1} \fr{1}{(\bx-\bx^\pp)^2 + \eps^2}  ,         
               \nonumber \\
    \lang \phi(\bx) \P_\chi(\bx^\pp) \rang &=&
             -\fr{1}{8\pi^2} \fr{1}{(\bx-\bx^\pp)^2 + \eps^2} ,
              \nonumber \\
    \lang \chi(\bx) \P_\chi(\bx^\pp) \rang &=& \lang \phi(\bx) \P_\phi(\bx^\pp) \rang
            = i\fr{1}{2\pi^2} \fr{\eps}{[(\bx-\bx^\pp)^2 +\eps^2]^2} ,
              \nonumber \\
    \lang \chi(\bx) \P_\phi(\bx^\pp) \rang  &=& 
       - \fr{1}{4\pi^2} \fr{(\bx-\bx^\pp)^2 - 3\eps^2}{[(\bx-\bx^\pp)^2 + \eps^2]^3},           
              \nonumber \\
    \lang \P_\chi(\bx) \P_\chi(\bx^\pp) \rang &=& 
       \fr{3b_1}{16\pi^4} \fr{(\bx-\bx^\pp)^2 - 3\eps^2}{[(\bx-\bx^\pp)^2 + \eps^2]^3},           
              \nonumber \\
    \lang \P_\phi(\bx) \P_\phi(\bx^\pp) \rang &=& 
         \fr{9b_1}{4\pi^4} \fr{(\bx-\bx^\pp)^4 -10\eps^2(\bx-\bx^\pp)^2 + 5\eps^4}{[(\bx-\bx^\pp)^2 + \eps^2]^5} ,
\end{eqnarray}
where $h$ is the real function defined in (\ref{function e_j and h}).

Here, only two cases $\lang \chi(\bx) \P_\chi(\bx^\pp) \rang$ and $\lang \phi(\bx) \P_\phi(\bx^\pp) \rang$ becomes complex, and thus only the equal-time commutation relations between $\chi$ and $\P_\chi$ and between $\phi$ and $\P_\phi$ become nontrivial as required.

Furthermore, we write down the cases with spatial derivatives,
\begin{eqnarray}
    \lang \pd_j \phi(\bx) \phi(\bx^\pp) \rang &=& 
          -\fr{1}{2b_1} \fr{(x-x^\pp)_j}{(\bx-\bx^\pp)^2} \left[ 1 -h(\bx-\bx^\pp) \right] ,
            \nonumber \\
    \lang \pd_j \phi(\bx) \pd_k \phi(\bx^\pp) \rang &=& 
          \fr{1}{2b_1} \biggl\{ \fr{(x-x^\pp)_j(x-x^\pp)_k}{(\bx-\bx^\pp)^2[(\bx-\bx^\pp)^2 +\eps^2]}
              \nonumber \\
         && + \left( \dl_{jk} - \fr{3(x-x^\pp)_j(x-x^\pp)_k}{(\bx-\bx^\pp)^2} \right) 
                \fr{1 -h(\bx-\bx^\pp)}{(\bx-\bx^\pp)^2}  \biggr\} ,
            \nonumber \\
    \lang \pd_j \phi(\bx) \pd^j \phi(\bx^\pp) \rang &=& 
          \fr{1}{2b_1} \fr{1}{(\bx-\bx^\pp)^2 +\eps^2} ,
              \nonumber \\
    \lang \lap3 \phi(\bx) \phi(\bx^\pp) \rang &=& 
          -\fr{1}{2b_1} \fr{1}{(\bx-\bx^\pp)^2 +\eps^2} ,
            \nonumber \\
    \lang \lap3 \phi(\bx) \pd_j \phi(\bx^\pp) \rang &=& \lang \chi(\bx) \pd_j \chi(\bx^\pp) \rang
          = -\fr{1}{b_1} \fr{(x-x^\pp)_j}{[(\bx-\bx^\pp)^2 +\eps^2]^2} ,
            \nonumber \\
    \lang \P_\chi(\bx) \pd_j \phi(\bx^\pp) \rang &=&
          -\fr{1}{4\pi^2} \fr{(x-x^\pp)_j}{[(\bx-\bx^\pp)^2 +\eps^2]^2} ,
            \nonumber \\
    \lang \P_\phi(\bx) \pd_j \phi(\bx^\pp) \rang &=& \lang \P_\chi(\bx) \pd_j \chi(\bx^\pp) \rang
         = -i \fr{2}{\pi^2} \fr{\eps(x-x^\pp)_j}{[(\bx-\bx^\pp)^2 +\eps^2]^3} ,
            \nonumber \\
    \lang \chi(\bx) \lap3 \chi(\bx^\pp) \rang &=&  
          -\fr{1}{b_1} \fr{(\bx-\bx^\pp)^2 -3 \eps^2}{[(\bx-\bx^\pp)^2 + \eps^2]^3}  ,         
               \nonumber \\
    \lang \pd_j \chi(\bx) \pd_k \chi(\bx^\pp) \rang &=&  
      -\fr{1}{b_1} \fr{\dl_{jk}[(\bx-\bx^\pp)^2 + \eps^2 ] -4 (x-x^\pp)_j (x-x^\pp)_k}{[(\bx-\bx^\pp)^2 + \eps^2]^3} ,
               \nonumber \\
    \lang \P_\chi(\bx) \lap3 \phi(\bx^\pp) \rang &=&  
          -\fr{1}{4\pi^2} \fr{(\bx-\bx^\pp)^2 -3 \eps^2}{[(\bx-\bx^\pp)^2 + \eps^2]^3}  ,         
               \nonumber \\
    \lang \P_\phi(\bx) \lap3 \phi(\bx^\pp) \rang &=& \lang \P_\chi(\bx) \lap3 \chi(\bx^\pp) \rang
            = - i\fr{6}{\pi^2} \fr{\eps [(\bx-\bx^\pp)^2 -\eps^2]}{[(\bx-\bx^\pp)^2 + \eps^2]^4} ,
              \nonumber \\
    \lang \P_\phi(\bx) \pd_j \chi(\bx^\pp) \rang &=& 
       - \fr{1}{\pi^2} \fr{(x-x^\pp)_j[(\bx-\bx^\pp)^2 - 5\eps^2]}{[(\bx-\bx^\pp)^2 + \eps^2]^4},           
              \nonumber \\
    \lang \pd_j \chi(\bx) \lap3 \chi(\bx^\pp) \rang &=& 
        \fr{4}{b_1} \fr{(x-x^\pp)_j[(\bx-\bx^\pp)^2 - 5\eps^2]}{[(\bx-\bx^\pp)^2 + \eps^2]^4} .         
\end{eqnarray}
Besides, the spatial derivatives of $\lang \phi(\bx) \chi(\bx^\pp) \rang$ and $\lang \P_\chi(\bx) \P_\phi(\bx^\pp) \rang$ vanish.

\section{Commutation Relations with $K_\mu$}
\setcounter{equation}{0}
\noindent

We here summarize the commutation relations between $K_\mu$ of special conformal transformations and various local operators in the Riegert field sector.

The commutation relations with $\calA$ and $\calB_j$ are given by
\begin{eqnarray}
    \left[ K_0, \calA \right] &=& i \biggl\{
        - \left( \eta^2 + \bx^2 \right) \pd_\eta \calA  
        - 2 \eta x^k \pd_k \calA  - 8 \eta \calA  - 2 x^k \calB_k 
            \nonumber \\
       && - 2 x^k \left( \l: \P_\chi \pd_k \chi \r: - \l: \pd_k \P_\chi \chi \r: \right)
            \nonumber \\
       && - 2x^k \fr{b_1}{4\pi^2} \left( \l: \lap3 \chi \pd_k \phi \r: + 2 \l: \pd_k \chi \lap3 \phi \r:
                                       - \l: \chi \pd_k \lap3 \phi \r: \right)
            \nonumber \\
       && - 2 \l: \P_\chi \chi \r: - \fr{b_1}{2\pi^2} \l: \chi \lap3 \phi \r: 
           -2 \P_\phi  - \fr{b_1}{2\pi^2} \lap3 \chi 
                     \biggr\},
            \nonumber \\
    \left[ K_j, \calA \right] &=& i \biggl\{
        -  \left( -\eta^2 + \bx^2 \right) \pd_j \calA  +2 x_j x^k \pd_k \calA 
        + 2\eta x_j \pd_\eta \calA 
             \nonumber \\
       && + 8 x_j \calA + 2 \eta \calB_j
          + 2\eta \left( \l: \P_\chi \pd_j \chi \r: - \l: \pd_j \P_\chi \chi \r: \right)
            \nonumber \\        
       &&  + 2 \eta \fr{b_1}{4\pi^2} \left( 
            \l: \lap3 \chi \pd_j \phi \r: + 2 \l: \pd_j \chi \lap3 \phi \r: - \l: \chi \pd_j \lap3 \phi \r:
               \right)
            \nonumber \\
       &&  + 2 \left( \l: \P_\chi \pd_j \phi \r: 
                      - \fr{b_1}{4\pi^2} \l: \lap3 \phi \pd_j \phi \r: \right)
           + \fr{b_1}{2\pi^2} \pd_j \l: \chi^2 \r:  
              \biggr\} ,
            \nonumber \\
    \left[ K_0, \calB_j \right] &=& i \biggl\{
        - \left( \eta^2 + \bx^2 \right) \pd_\eta \calB_j - 2 \eta x^k \pd_k \calB_j 
        - 8 \eta \calB_j 
             \nonumber \\
        && - 2 x_j \left( -\fr{4\pi^2}{b_1} \l: \P_\chi^2 \r: + \l: \P_\phi \chi \r: \right)
           -2 x^k \pd_k \l: \P_\chi \pd_j \phi \r:
              \nonumber \\
        &&  + 2 x^k \fr{b_1}{2\pi^2} \left( \l: \pd_k \chi \pd_j \chi \r: 
                     + \l: \pd_k \lap3 \phi \pd_j \phi \r: \right)
               \nonumber \\
        &&  - 8 \left( \l: \P_\chi \pd_j \phi \r: 
                       - \fr{b_1}{4\pi^2} \l: \lap3 \phi \pd_j \phi \r: \right) 
            + \fr{b_1}{4\pi^2} \pd_j \l: \chi^2 \r:
              \biggr\},
            \nonumber \\
    \left[ K_j, \calB_k \right] &=& i \biggl\{
        - \left( -\eta^2 + \bx^2 \right) \pd_j \calB_k  + 2 x_j x^l \pd_l \calB_k 
        + 2\eta x_j \pd_\eta \calB_k  
              \nonumber \\
        && + 8 x_j \calB_k  -2 x_k \calB_j  + 2\dl_{jk} x^l \calB_l
              \nonumber \\
        && + 2\eta \left( \pd_j \l: \P_\chi \pd_k \phi \r:
             - \fr{b_1}{2\pi^2} \left[ \l: \pd_j \chi \pd_k \chi \r: 
                    + \l: \pd_j \lap3 \phi \pd_k \phi \r: \right] \right) 
              \nonumber \\ 
        && + 2 \eta \dl_{jk} \left( -\fr{4\pi^2}{b_1} \l: \P_\chi^2 \r: + \l: \P_\phi \chi \r: \right)
           + 2 \dl_{jk} \l: \P_\chi \chi \r:
              \nonumber \\
        && + \fr{b_1}{2\pi^2} \left( \l: \pd_j \chi \pd_k \phi \r: - \l: \pd_k \chi \pd_j \phi \r: \right)
           + 2 \dl_{jk} \P_\phi 
               \biggr\} 
\end{eqnarray}
and the commutation relations with the other local operators that appear in the generators are given by
\begin{eqnarray}
    \left[ K_0, \l: \P_\chi \pd_j \phi \r: \right] &=& i \biggl\{
        -  \left( \eta^2 + \bx^2 \right) \pd_\eta \l: \P_\chi \pd_j \phi \r: 
        - 2 \eta x^k \pd_k \l: \P_\chi \pd_j \phi \r:  - 6 \eta \l: \P_\chi \pd_j \phi \r:
            \nonumber \\
       &&  - 2 x_j \l: \P_\chi \chi \r: 
          + 2 x^k \fr{b_1}{2\pi^2} \l: \pd_k \chi \pd_j \phi \r: 
          + \fr{b_1}{2\pi^2} \l: \chi \pd_j \phi \r:   
              \biggr\},
            \nonumber \\
    \left[ K_j, \l: \P_\chi \pd_k \phi \r: \right] &=& i \biggl\{
        -  \left( -\eta^2 + \bx^2 \right) \pd_j \l: \P_\chi \pd_k \phi \r: 
        +2 x_j x^l \pd_l \l: \P_\chi \pd_k \phi \r:   
              \nonumber \\
       && + 2 \eta x_j \pd_\eta \l: \P_\chi \pd_k \phi \r: 
          + 6 x_j \l: \P_\chi \pd_k \phi \r: - 2 x_k \l: \P_\chi \pd_j \phi \r:
            \nonumber \\
       &&  + 2 \dl_{jk} x^l \l: \P_\chi \pd_l \phi \r: 
           + 2 \eta \left( \dl_{jk} \l: \P_\chi \chi \r: 
                           - \fr{b_1}{2\pi^2} \l: \pd_j \chi \pd_k \phi \r: \right)
             \nonumber \\
       && - \fr{b_1}{2\pi^2} \l: \pd_j \phi \pd_k \phi \r: + 2\dl_{jk} \P_\chi 
              \biggr\},
           \nonumber \\
    \left[ K_0, \l: \P_\chi \chi \r: \right] &=& i \biggl\{
        -  \left( \eta^2 + \bx^2 \right) \pd_\eta \l: \P_\chi \chi \r: 
        - 2 \eta x^k \pd_k \l: \P_\chi \chi \r:  - 6 \eta \l: \P_\chi \chi \r:  
           \nonumber \\
       && - 2 x^k \l: \P_\chi \pd_k \phi \r: 
          + 2 x^k \fr{b_1}{4\pi^2} \pd_k \l: \chi^2 \r: 
          + \fr{b_1}{2\pi^2} \l: \chi^2 \r:    - 2 \P_\chi
              \biggr\},
            \nonumber \\
    \left[ K_j, \l: \P_\chi \chi \r: \right] &=& i \Bigl\{
         -  \left( -\eta^2 + \bx^2 \right) \pd_j \l: \P_\chi \chi \r: 
         + 2 x_j x^k \pd_k \l: \P_\chi \chi \r:  
            \nonumber \\
       && + 2\eta x_j \pd_\eta \l: \P_\chi \chi \r:  + 6 x_j \l: \P_\chi \chi \r: 
            \nonumber \\
       && + 2 \eta \left( \l: \P_\chi \pd_j \phi \r: - \fr{b_1}{4\pi^2} \pd_j \l: \chi^2 \r: \right)
          - \fr{b_1}{2\pi^2} \l: \chi \pd_j \phi \l:
              \biggr\} ,
            \nonumber \\
    \left[ K_0, \l: \chi^2 \r: \right] &=& i \biggl\{
        -  \left( \eta^2 + \bx^2 \right) \pd_\eta \l: \chi^2 \r: 
        - 2 \eta x^k \pd_k \l: \chi^2 \r: - 4 \eta \l: \chi^2 \r: 
           \nonumber \\
       &&  - 4 x^k \l: \chi \pd_k \phi \r: - 4 \chi
              \biggr\},
            \nonumber \\
    \left[ K_j, \l: \chi^2 \r: \right] &=& i \biggl\{
         -  \left( -\eta^2 + \bx^2 \right) \pd_j \l: \chi^2 \r: 
         + 2 x_j x^k \pd_k \l: \chi^2 \r: 
            \nonumber \\
       &&  + 2\eta x_j \pd_\eta \l: \chi^2 \r:  + 4 x_j \l: \chi^2 \r: 
           + 4 \eta \l: \chi \pd_j \phi \r: 
              \biggr\} ,
            \nonumber \\
    \left[ K_0, \l: \pd_k \phi \pd^k \phi \r: \right] &=& i \biggl\{
        -  \left( \eta^2 + \bx^2 \right) \pd_\eta \l: \pd_k \phi \pd^k \phi \r: 
        - 2 \eta x^l \pd_l \l: \pd_k \phi \pd^k \phi \r:  
           \nonumber \\
       && - 4 \eta \l: \pd_k \phi \pd^k \phi \r:  - 4 x^k \l: \chi \pd_k \phi \r:
              \biggr\},
            \nonumber \\
    \left[ K_j, \l: \pd_k \phi \pd^k \phi \r: \right] &=& i \biggl\{
         -  \left( -\eta^2 + \bx^2 \right) \pd_j \l: \pd_k \phi \pd^k \phi \r: 
         + 2 x_j x^l \pd_l \l: \pd_k \phi \pd^k \phi \r: 
            \nonumber \\
       &&  + 2\eta x_j \pd_\eta \l: \pd_k \phi \pd^k \phi \r:  + 4 x_j \l: \pd_k \phi \pd^k \phi \r: 
           + 4 \eta \l: \chi \pd_j \phi \r: 
           + 4 \pd_j \phi
              \biggr\} ,
            \nonumber \\
    \left[ K_0, \l: \chi \pd_j \phi \r: \right] &=& i \biggl\{
        -  \left( \eta^2 + \bx^2 \right) \pd_\eta \l: \chi \pd_j \phi \r: 
        - 2 \eta x^k \pd_k \l: \chi \pd_j \phi \r:  
           \nonumber \\
       && - 4 \eta \l: \chi \pd_j \phi \r:  - 2 x^k \l: \pd_k \phi \pd_j \phi \r:
              \biggr\},
            \nonumber \\
    \left[ K_j, \l: \chi \pd_k \phi \r: \right] &=& i \biggl\{
         -  \left( -\eta^2 + \bx^2 \right) \pd_j \l: \chi \pd_k \phi \r: 
         + 2 x_j x^l \pd_l \l: \chi \pd_k \phi \r: 
            \nonumber \\
       &&  + 2\eta x_j \pd_\eta \l: \chi \pd_k \phi \r:  + 4 x_j \l: \chi \pd_k \phi \r:
           - 2 x_k \l: \chi \pd_j \phi \r: 
              \nonumber \\
       &&  + 2 \dl_{jk} x^l \l: \chi \pd_l \phi \r: 
           + 2 \eta \left( \dl_{jk} \l: \chi^2 \r: + \l: \pd_j \phi \pd_k \phi \r: \right) 
           + 2 \dl_{jk} \chi 
              \biggr\} .
\end{eqnarray}
Here, all quantum correction terms disappear.

\section{Stress Tensors of Traceless Tensor Fields}
\setcounter{equation}{0}
\noindent

The square of the Weyl tensor is defined by 
\begin{eqnarray}
   C^2_{\mu\nu\lam\s} = R^2_{\mu\nu\lam\s} - 2 R^2_{\mu\nu} + \fr{1}{3} R^2 .
\end{eqnarray}

Since the Weyl action is conformally invariant, there is no dependence on the conformal factor of the metric field, or the Riegert field. Therefore, we here use the metric field without the conformal factor defined by
\begin{eqnarray}
     \bg_{\mu\nu} = (e^h )_{\mu\nu} = \eta_{\mu\nu} + h_{\mu\nu} + \half h_{\mu\lam}h^\lam_{~\nu} + \cdots .
\end{eqnarray}
We also disregard the coupling constant $t$ for simplicity.

The stress tensor of the traceless tensor field derived from the Weyl action is given by the Bach tensor,
\begin{eqnarray}
    {\bar T}_{\mu\nu} &=& 4 \bnb^2 \bR_{\mu\nu} - \fr{4}{3} \bnb_\mu \bnb_\nu \bR 
                       + 8 \bR^{\lam\s}\bR_{\mu\lam\nu\s} - \fr{8}{3}\bR \bR_{\mu\nu}
                  \nonumber \\
             && - \bg_{\mu\nu}  \left( \fr{2}{3} \bnb^2 \bR + 2 \bR^{\lam\s} \bR_{\lam\s} 
                                      -\fr{2}{3} \bR^2 \right)  ,
\end{eqnarray}
where the curvatures with the bar on them are defined in terms of $\bg_{\mu\nu}$. It is expanded in the field as
\begin{eqnarray}
     {\bar T}_{\mu\nu} = {\bar T}^{(1)}_{\mu\nu} + {\bar T}^{(2)}_{\mu\nu} + \cdots .
\end{eqnarray}
The linear term is given by
\begin{equation}
    {\bar T}^{(1)}_{\mu\nu} = - 2 \pd^4 h_{\mu\nu} + 4 \pd^2 \pd_{(\mu} \chi_{\nu)} 
         - \fr{4}{3} \pd_\mu \pd_\nu \pd_\lam \chi^\lam - \fr{2}{3} \eta_{\mu\nu} \pd^2 \pd_\lam \chi^\lam ,
\end{equation}
where $\chi_\nu = \pd_\lam h^\lam_{~\nu}$. The equations of motion are represented by ${\bar T}^{(1)}_{\mu\nu} =0$.

The generator of conformal symmetry defined by (\ref{definition of Q_zeta}) is derived from the bilinear term. We here consider the following combination of the stress tensor: 
\begin{eqnarray}
    T_{\mu\nu}  = {\bar T}^{(2)}_{\mu\nu} - h_{(\mu}^\lam {\bar T}^{(1)}_{\nu)\lam} ,
\end{eqnarray}
which is equivalent to ${\bar T}^{(2)}_{\mu\nu}$ up to the equations of motion. The indices are then contracted by the flat background metric so that the trace $\bg^{\mu\nu}{\bar T}_{\mu\nu}$ can be expressed as $\eta^{\mu\nu}T_{\mu\nu}$.  It is given by 
\begin{eqnarray}
  && T_{\mu\nu}
     = -6 \pd^2 h_{\lam\s} \pd^\lam \pd_{(\mu} h^\s_{\nu)}  +4 \pd_\rho \pd_\lam h_{\s(\mu} \pd_{\nu)} \pd^\rho h^{\lam\s}
       +4 \pd^2 h_{\lam\s} \pd^\lam \pd^\s h_{\mu\nu} 
                \nonumber \\
   && \quad 
       -\fr{4}{3} \pd_\rho \pd_\mu h_{\lam\s} \pd^\rho \pd_\nu h^{\lam\s} 
       -4 \pd_\rho \pd_\lam h^\s_\mu \pd^\rho \pd_\s h^\lam_\nu   -4 \pd^2 h_{\lam(\mu} \pd^\lam \chi_{\nu)}
                \nonumber \\
   && \quad 
       -\fr{2}{3} \pd_\mu \chi^\lam \pd_\nu \chi_\lam     +2 \pd_{(\mu} \chi^\lam \pd^2 h_{\nu)\lam}
       +2 \pd^\lam \chi_\mu \pd_\lam \chi_\nu    -\fr{8}{3} \pd_\mu \pd_\nu h_{\lam\s} \pd^\lam \chi^\s
                \nonumber \\
   && \quad 
       +4 \pd_\lam \chi_\s \pd^\s \pd_{(\mu} h^\lam_{\nu)}    +2 \pd^2 h_{\lam\s} \pd_\mu \pd_\nu h^{\lam\s}
       -\fr{8}{3} \pd_\lam \chi^\lam \pd_{(\mu} \chi_{\nu)}  +\fr{4}{3} \pd_\lam \chi^\lam \pd^2 h_{\mu\nu}
               \nonumber \\
   && \quad 
       -2 \pd_{(\mu} h^\lam_{\nu)} \pd^2 \chi_\lam   -2 \chi_\lam \pd^2 \pd_{(\mu} h^\lam_{\nu)} 
       -4 \pd_\rho h_{\lam\s} \pd^\rho \pd^\lam \pd_{(\mu} h^\s_{\nu)}   
       +6 \pd^2 \pd^\lam h^\s_{(\mu} \pd_{\nu)} h_{\lam\s} 
               \nonumber \\
   && \quad 
       +2 \pd_\lam h_{\s(\mu} \pd^2 \pd_{\nu)} h^{\lam\s}    +2 \pd_\lam h_{\mu\nu} \pd^2 \chi^\lam
       +4 \chi_\lam \pd^2 \pd^\lam h_{\mu\nu}      +4 \pd_\rho h_{\lam\s} \pd^\rho \pd^\lam \pd^\s h_{\mu\nu}
               \nonumber \\
   && \quad 
       -2 \pd_{(\mu} h^{\lam\s} \pd^2 \pd_{\nu)} h_{\lam\s}    -8 \pd_\s h_{\lam(\mu} \pd^2 \pd^\lam h^\s_{\nu)}
       -4 \chi^\lam \pd_\lam \pd_{(\mu} \chi_{\nu)}       -4 \pd^\lam \pd^\s \chi_{(\mu} \pd_{\nu)} h_{\lam\s}
               \nonumber \\
   && \quad 
       -\fr{4}{3} \pd_{(\mu} h^{\lam\s} \pd_{\nu)} \pd_\lam \chi_\s   +4 \pd_\lam h_{\s(\mu} \pd_{\nu)} \pd^\s \chi^\lam
       +\fr{4}{3} \chi_\lam \pd_\mu \pd_\nu \chi^\lam 
               \nonumber \\
   && \quad 
       +\fr{2}{3} \pd_\rho h_{\lam\s} \pd_\mu \pd_\nu \pd^\rho h^{\lam\s}  
       +\fr{4}{3} \pd_{(\mu} h^\lam_{\nu)} \pd_\lam \pd_\s \chi^\s     
       -\fr{2}{3} \pd^\lam h_{\mu\nu} \pd_\lam \pd_\s \chi^\s
               \nonumber \\
   && \quad 
       -2 h_{\lam\s} \pd^2 \pd^\lam \pd_{(\mu} h^\s_{\nu)}    +4 h_{\lam\s} \pd^2 \pd^\lam \pd^\s h_{\mu\nu}
       -4 h_{\lam\s} \pd^\lam \pd^\s \pd_{(\mu} \chi_{\nu)}  
               \nonumber \\
   && \quad 
       +\fr{4}{3} h_{\lam\s} \pd_\mu \pd_\nu \pd^\lam \chi^\s   -2 h_{\lam(\mu} \pd^2 \pd^\lam \chi_{\nu)}
       +\fr{4}{3} h_{\lam(\mu} \pd_{\nu)} \pd^\lam \pd_\s \chi^\s
               \nonumber \\
   && \quad 
       - \eta_{\mu\nu} \biggl(
       \fr{1}{3} \pd_\lam \chi_\s \pd^\lam \chi^\s   -\fr{8}{3} \pd^2 h_{\lam\s} \pd^\lam \chi^\s
       -\fr{1}{3} \pd_\kappa \pd_\rho h_{\lam\s} \pd^\kappa \pd^\rho h^{\lam\s}
       + \pd_\lam \chi_\s \pd^\s \chi^\lam
               \nonumber \\
   && \quad 
       +\fr{1}{2} \pd^2 h_{\lam\s} \pd^2 h^{\lam\s}    -\fr{2}{3} \pd_\lam \chi^\lam \pd_\s \chi^\s
       -\fr{2}{3} \chi_\lam \pd^2 \chi^\lam    -\fr{4}{3} \pd_\rho h_{\lam\s} \pd^\rho \pd^\lam \chi^\s
               \nonumber \\
   && \quad 
       -\fr{1}{3} \pd_\rho h_{\lam\s} \pd^2 \pd^\rho h^{\lam\s}    -\fr{2}{3} \chi^\lam \pd_\lam \pd_\s \chi^\s
       -\fr{2}{3} h_{\lam\s} \pd^2 \pd^\lam \chi^\s    -\fr{2}{3} h_{\lam\s} \pd^\lam \pd^\s \pd_\rho \chi^\rho
         \biggr) .
              \nonumber \\
\end{eqnarray}

Here, the time-derivatives of various field variables in the radiation gauge are expressed as $\pd_\eta \h^{ij} = \u^{ij}$, $\pd_\eta^2 \h^{ij} = - \P_\u^{ij}$, $\pd_\eta^3 \h^{ij} = \P_\h^{ij} + 2 \lap3 \u^{ij}$ and $\pd_\eta^4 \h^{ij} = -2 \lap3 \P_\u^{ij} - \dlap3 \h^{ij}$ for the transverse traceless tensor mode and $\pd_\eta \h^j = \invlap3 \P^j/2$, $\pd_\eta^2 \h^j = \lap3 \h^j$, $\pd_\eta^3 \h^j = \P^j/2$ and $\pd_\eta^4 \h^j = \dlap3 \h^j$ for the transverse vector mode.

Using these equations, we write down the stress tensor in terms of six canonical variables as follows. 
The time-time components denoted by $T_{00} = T^{(t)}_{00} + T^{(v)}_{00} + T^{(c)}_{00}$ are given by
\begin{eqnarray}
   T^{(t)}_{00} 
   &=& -\half \P^{ij}_\u \P_{ij}^\u                     + \P^{ij}_\h \u_{ij}     
       -\P^{ij}_\u \lap3 \h_{ij}                         - \pd^k \P^{ij}_\u \pd_k \h_{ij}   
       + \fr{1}{3} \u^{ij} \lap3 \u_{ij}          
             \nonumber \\
   &&  - \fr{2}{3} \pd^k \u^{ij} \pd_k \u_{ij}             + \half \lap3 \h^{ij} \lap3 \h_{ij}
       - \fr{1}{3} \pd^k \pd^l \h^{ij} \pd_k \pd_l \h_{ij}       - \fr{1}{3} \pd^k \h^{ij} \pd_k \lap3 \h_{ij} ,
             \nonumber \\
  T^{(v)}_{00}
   &=&  - \fr{1}{3} \P^k \invlap3 \P_k                  - \fr{7}{12} \invlap3 \pd^k \P^l \invlap3 \pd_k \P_l
       + \fr{1}{4} \invlap3 \pd^k \P^l \invlap3 \pd_l \P_k
             \nonumber \\ 
   &&  - \fr{5}{3} \lap3 \h^k \lap3 \h_k                  + \fr{2}{3} \pd^m \pd^l \h^k \pd_m \pd_l \h_k
       + \fr{4}{3} \pd^l \h^k \pd_l \lap3 \h_k            + \fr{4}{3} \h^k \dlap3 h_k
             \nonumber \\
   &&  - 4 \pd^m \pd^l \h^k \pd_m \pd_k \h_l              - 4 \pd^l \h^k \pd_k \lap3 \h_l ,
             \nonumber \\
  T^{(c)}_{00}
   &=&  - \P^{ij}_\u \invlap3 \pd_i \P_j                 - 2 \P^{ij}_\h \pd_i \h_j
       - \fr{2}{3} \h^{ij} \pd_i \P_j                     - \fr{5}{3} \lap3 \h^{ij} \invlap3 \pd_i \P_j
             \nonumber \\
   &&  - \fr{4}{3} \pd^k \h^{ij} \invlap3 \pd_k \pd_i \P_j     - 2 \lap3 \u^{ij} \pd_i \h_j
       + 4 \u^{ij} \pd_i \lap3 \h_j                            + 4 \pd^k \u^{ij} \pd_k \pd_i \h_j 
             \nonumber \\
\end{eqnarray}
and the $(0j)$ components denoted by $T_{0j} = T^{(t)}_{0j} + T^{(v)}_{0j} + T^{(c)}_{0j}$ are given by
\begin{eqnarray}
  T^{(t)}_{0j}
   &=& \P^{kl}_\h \pd_j \h_{kl}                         - \P^{kl}_\h \pd_k \h_{lj}
       + \pd^k \P^l_{\h j} \h_{kl}                       + \fr{2}{3} \P^{kl}_\u \pd_j \u_{kl}
       - \fr{1}{3} \pd_j \P^{kl}_\u \u_{kl}                               
             \nonumber \\
   &&  - \P^{kl}_\u \pd_k \u_{lj}                        + \pd^k \P^l_{\u j} \u_{kl}
       + 2 \pd_j \u^{kl} \lap3 \h_{kl}                    - \fr{4}{3} \pd^m \u^{kl} \pd_j \pd_m \h_{kl}
             \nonumber \\
   &&  - \u^{kl} \pd_j \lap3 \h_{kl}                      + \lap3 \u^{kl} \pd_j \h_{kl}
       + \fr{2}{3} \pd_j \pd^m \u^{kl} \pd_m \h_{kl}      - 3 \pd^k \u^l_{~j} \lap3 \h_{kl}
             \nonumber \\
   &&  - 2 \pd^m \pd^k \u^l_{~j} \pd_m \h_{kl}               + \lap3 \pd^k \u^l_{~j} \h_{kl}
       + 2 \pd^m \u^{kl} \pd_m \pd_k \h_{lj}              + 3 \u^{kl} \pd_k \lap3 \h_{lj}                      
             \nonumber \\
   &&  - \lap3 \u^{kl} \pd_k \h_{lj} ,
                 \nonumber \\
  T^{(v)}_{0j}
   &=&  - \fr{1}{3} \P^k \pd_j \h_k                       + \fr{2}{3} \pd_j \P^k \h_k
       - 2 \invlap3 \pd_j \P^k \lap3 \h_k                + \fr{1}{3} \invlap3 \pd^l \P^k \pd_j \pd_l \h_k
            \nonumber \\
   &&  + \fr{1}{3} \invlap3 \pd_j \pd^l \P^k \pd_l \h_k     + \P^k \pd_k \h_j
       + \invlap3 \pd^l \P^k \pd_l \pd_k \h_j               - 2 \pd^k \P_j \h_k 
                          \nonumber \\
   &&  - 3 \invlap3 \pd^l \pd^k \P_j \pd_l \h_k             - \invlap3 \pd^l \P^k \pd_j \pd_k \h_l
       - \invlap3 \pd_j \pd^l \P^k \pd_k \h_l ,
            \nonumber \\
  T^{(c)}_{0j} 
   &=& 4 \P^{kl}_\u \pd_k \pd_l \h_j                        - 3 \P^k_{\u j} \lap3 \h_k
       - 2 \pd^k \P^l_{\u j} \pd_k \h_l                     + \lap3 \P^k_{\u j} \h_k
       - 4 \pd^k \P^l_{\u j} \pd_l \h_k 
            \nonumber \\
   &&  + \pd_j \P^{kl}_\u \pd_k \h_l                        - 3 \P^{kl}_\u \pd_j \pd_k \h_l  
       - \lap3 \u^k_{~j} \invlap3 \P_k                      - \pd^k \u^l_{~j} \invlap3 \pd_k \P_l
            \nonumber \\
   &&  + \pd^k \u^l_{~j} \invlap3 \pd_l \P_k                + \fr{1}{3} \pd_j \u^{kl} \invlap3 \pd_k \P_l 
       + \fr{4}{3} \u^{kl} \invlap3 \pd_j \pd_k \P_l         - \u^{kl} \invlap3 \pd_k \pd_l \P_j 
            \nonumber \\
   &&  - \lap3 \h^k_{~j} \lap3 \h_k                          + \dlap3 \h^k_{~j} \h_k
       - 3 \lap3 \h^{kl} \pd_j \pd_k \h_l                    + 2 \pd_j \pd^m \h^{kl} \pd_m \pd_k \h_l
            \nonumber \\
   &&  + \pd_j \lap3 \h^{kl} \pd_k \h_l                      - 2 \pd^m \h^{kl} \pd_j \pd_m \pd_k \h_l
       + \fr{2}{3} \pd_j \h^{kl} \pd_k \lap3 \h_l            - \fr{4}{3} \h^{kl} \pd_j \pd_k \lap3 \h_l
            \nonumber \\
   &&  + 4 \lap3 \h^{kl} \pd_k \pd_l \h_j                   + 4 \pd^m \h^{kl} \pd_m \pd_k \pd_l \h_j        
       + 2 \h^{kl} \pd_k \pd_l \lap3 \h_j                   - 4 \pd^m \pd^k \h^l_{~j} \pd_m \pd_l \h_k
            \nonumber \\
   &&  - 4 \lap3 \pd^k \h^l_{~j} \pd_l \h_k
       - 2 \pd^k \h^l_{~j} \pd_l \lap3 \h_k  .                 
\end{eqnarray}

\section{Conformal Algebra for Scalar Fields}
\setcounter{equation}{0}
\noindent

As a simple exercise, we here present computations of conformal algebra for the case of real scalar field conformally coupled to gravity. The action is given by 
\begin{equation}
     I = -\half \int d^4 x \sq{-\hg} \left( \hg^{\mu\nu} \pd_\mu X \pd_\nu X + \fr{1}{6} \hR X^2 \right) ,
          \label{scalar action}
\end{equation}
where $\hg_{\mu\nu}$ is the background metric taken to be flat. The canonical commutation relation is defined by $[ X(\eta, \bx), \P_X(\eta, \bx^\pp) ] = i \dl_3 (\bx -\bx^\pp)$, where $\P_X= \pd_\eta X$ is the conjugate momentum. The annihilation part of $X$ is expanded as 
\begin{eqnarray}
    X_<(x) = \int \fr{d^3 \bk}{(2\pi)^{3/2}} \fr{1}{\sq{2\om}} \vphi(\bk) e^{ik_\mu x^\mu}
\end{eqnarray}
with the commutation relation $[ \vphi(\bk), \vphi^\dag(\bk^\pp)] = \dl_3 (\bk -\bk^\pp)$.

The stress tensor is given by
\begin{eqnarray}
    T_{\mu\nu} = \fr{2}{3} \pd_\mu X \pd_\nu X - \fr{1}{3} X \pd_\mu \pd_\nu X 
                 - \fr{1}{6} \eta_{\mu\nu} \pd^\lam X \pd_\lam X .
\end{eqnarray}
The $(00)$ and $(0j)$ components are given by $T_{00} = \P_X^2/2 - X \lap3 X/3 + \pd_i X \pd^i X/6$ and $T_{0j} = 2 \P_X \pd_j X/3 - X \pd_j \P_X/3$, respectively.

The generators of conformal symmetry are expressed as follows:
\begin{eqnarray}
        P_0 &=& H = \int d^3 \bx \calA,  \qquad  P_j = \int d^3 \bx \calB_j, 
              \nonumber \\
     M_{0j} &=& \int d^3 \bx \left( -\eta \calB_j - x_j \calA \right),  \qquad
     M_{ij} = \int d^3 \bx \left(  x_i \calB_j -x_j \calB_i \right) ,
              \nonumber \\
     D &=& \int d^3 \bx \left( \eta \calA + x^k \calB_k  + \l: \P_X X \r: \right) ,
              \nonumber \\
     K_0 &=& \int d^3 \bx \left\{ 
               \left( \eta^2 + \bx^2 \right) \calA + 2\eta x^k \calB_k  
               + 2\eta \l: \P_X X \r: + \half \l: X^2 \r: \right\},
                 \nonumber \\
     K_j &=& \int d^3 \bx \left\{
               \left( -\eta^2 + \bx^2 \right) \calB_j  -2 x_j x^k \calB_k  
              - 2\eta x_j \calA  - 2x_j \l: \P_X X \r:  \right\} ,
\end{eqnarray}
where
\begin{eqnarray}
    \calA &=& \half \l: \P_X^2 \r: -\half \l: X \lap3 X \r: ,  \qquad    \calB_j = \l: \P_X \pd_j X \r: .             
\end{eqnarray}

The singular parts of OPEs for canonical field variables are computed at the equal time as
\begin{eqnarray}
      \lang X(\bx) X(\bx^\pp) \rang 
        &=& \fr{1}{4\pi^2} \fr{1}{(\bx -\bx^\pp)^2 + \eps^2}  ,
           \nonumber \\
      \lang X(\bx) \P_X(\bx^\pp) \rang 
        &=& i \fr{1}{2\pi^2} \fr{\eps}{[ (\bx -\bx^\pp)^2 + \eps^2]^2} ,
           \nonumber \\
      \lang \P_X(\bx) \P_X(\bx^\pp) \rang 
        &=& - \fr{1}{2\pi^2} \fr{(\bx -\bx^\pp)^2 - 3 \eps^2}{[ (\bx -\bx^\pp)^2 + \eps^2]^3} .
\end{eqnarray}
Using these expressions, the equal-time commutation relations between various local operators are computed as
\begin{eqnarray}
    \left[ \calA(\bx), \calA(\by) \right]
    &=& \half i \lap3_x \dl_3 (\bx-\by) \left( \l: \P_X(\bx)X(\by) \r: - \l: X(\bx)\P_X(\by) \r: \right),
            \nonumber \\
    \left[ \calB_j(\bx), \calB_k(\by) \right]
    &=& i \pd_k^x \dl_3 (\bx-\by) \l: \pd_j X(\bx)\P_X(\by) \r: 
        + i \pd_j^x \dl_3 (\bx-\by) \l: \P_X(\bx)\pd_k X(\by) \r:,  
          \nonumber \\
   \left[ \calA(\bx), \calB_j(\by) \right]
     &=& i \pd_j^x \dl_3(\bx-\by) \l: \P_X(\bx)\P_X(\by) \r: 
          - \half i \dl_3(\bx-\by) \l: \lap3 X \pd_j X(\by) \r:
             \nonumber \\
     && \quad 
          -\half i \lap3_x \dl_3(\bx-\by) \l: X(\bx)\pd_j X(\by) \r:
          - i \fr{2}{\pi^2} f_j(\bx-\by) 
\end{eqnarray}
and
\begin{eqnarray}
    \left[ \calA(\bx), :\P_X X(\by): \right] &=&
       -i \dl_3 (\bx-\bx) \left( \l: \P_X^2(\by) \r: + \half \l: X \lap3 X(\by) \r: \right) 
           \nonumber \\
        && - \half i \lap3_x \dl_3 (\bx-\by) \l: X(\bx)X(\by) \r: + i \fr{10}{\pi^2} f(\bx-\by) ,
           \nonumber \\
    \left[ \calB_j(\bx), \l: \P_X X(\by) \r: \right] &=&
       -i \dl_3 (\bx-\bx) \calB_j(\by) + i \pd_j^x \dl(\bx-\by) \l: \P_X(\bx)X(\by) \r: .
           \nonumber \\
\end{eqnarray}
Here, the quantum correction functions $f_j$ and $f$ are defined by (\ref{function f_j}) and (\ref{scalar function f}), respectively. Using these commutation relations, we find that all quantum corrections disappear and the conformal algebra is closed.

Next, we consider the transformation property of the composite operator $\l: X^n \r:$. The commutators between this operator and above operators are given by
\begin{eqnarray}
     \left[ \calA(\bx), \l: X^n(\by) \r: \right] 
      &=& -i \dl_3 (\bx -\by) \pd_\eta \l: X^n(\by) \r: ,
         \nonumber \\
      \left[ \calB_j(\bx), \l: X^n(\by) \r: \right] 
       &=& -i \dl_3 (\bx-\by ) \pd_j \l: X^n (\by) \r: 
              \nonumber \\
       &&   + i \fr{1}{2\pi^2}n(n-1) g_j(\bx-\by) \l: X^{n-2}(\by) \r: ,
            \nonumber \\
      \left[ \l: \P_X X(\bx) \r:, \l: X^n(\by) \r: \right] 
       &=& -i n \dl_3 (\bx-\by ) \l: X^n (\by) \r: 
              \nonumber \\
        &&  + i \fr{3}{2\pi^2}n(n-1) g(\bx-\by) \l: X^{n-2}(\by) \r: ,
\end{eqnarray}
where the functions $g_j$ and $g$ are defined by (\ref{function g_j}) and (\ref{scalar function g}), respectively.
The transformation laws of the composite operator $\l: X^n \r:$ are then given by
\begin{eqnarray}
   i\left[ P_\mu, \l: X^n(x) \r: \right] &=& \pd_\mu \l: X^n(x) \r: ,
             \nonumber \\
   i \left[ M_{\mu\nu}, \l: X^n(x) \r: \right] &=& \left( x_\mu \pd_\nu - x_\nu \pd_\mu \right) \l: X^n(x) \r: ,
             \nonumber \\
   i\left[ D, \l: X^n(x) \r: \right] &=& \left( x^\mu \pd_\mu + n \right) \l: X^n(x) \r: ,
             \nonumber \\
   i \left[ K_\mu, \l: X^n(x) \r: \right] 
       &=& \left( x^2 \pd_\mu - 2x_\mu x^\nu \pd_\nu - 2x_\mu n \right) \l: X^n(x) \r: .
        \label{transformation of X^n}
\end{eqnarray}
Here, all quantum correction terms cancel out and thus $\l: X^n \r:$ transforms as a conformal field with dimension $n$.

From (\ref{transformation of X^n}), we find that $\int d^4 x \! \l: X^4(x) \r:$ and $\int  d^4 x V_\b(x) \! \l: X^2(x) \r: $ with $h_\b =2$, for instance, become diffeomorphism invariant.


\end{document}